\documentclass[aps, prd, 10pt, notitlepage, superscriptaddress, nofootinbib,numbers,longbibliography]{revtex4-1}
\usepackage[utf8]{inputenc}
\usepackage[T1]{fontenc}
\usepackage{anyfontsize}

\usepackage{amsmath}
\usepackage{amssymb}
\usepackage{amsfonts}
\usepackage{mathrsfs}

\usepackage{microtype}

\usepackage{graphicx}
\usepackage{epsfig}
\usepackage{subcaption}

\usepackage[dvipsnames]{xcolor}
\usepackage{hyperref}
\hypersetup{
    colorlinks=true,
    citecolor=Purple,
    linkcolor=Purple,
    urlcolor=Purple,
    linktocpage=true,
    breaklinks=true
}

\usepackage{orcidlink}

\usepackage[capitalize]{cleveref}
\begin{document}

\title{Three dimensional black bounces in $f(R)$ gravity}

\author{Marcos V. de S. Silva}
\email{marcos.sousa@uva.es}
\affiliation{Department of Theoretical Physics, Atomic and Optics, Campus Miguel Delibes, \\ University of Valladolid UVA, Paseo Bel\'en, 7, 47011 - Valladolid, Spain}

\author{Manuel E. Rodrigues}
\email{esialg@gmail.com}
\affiliation{Faculdade de Ci\^{e}ncias Exatas e Tecnologia, 
Universidade Federal do Par\'{a}\\
Campus Universit\'{a}rio de Abaetetuba, 68440-000, Abaetetuba, Par\'{a}, Brazil}
\affiliation{Faculdade de F\'{\i}sica, Programa de P\'{o}s-Gradua\c{c}\~ao em 
F\'isica, Universidade Federal do 
Par\'{a}, 66075-110, Bel\'{e}m, Par\'{a}, Brazil}

\author{C. F. S. Pereira}
	\email{carlosfisica32@gmail.com}
	\affiliation{Departamento de F\'isica e Qu\'imica, Universidade Federal do Esp\'irito Santo, Av.Fernando Ferrari, 514, Goiabeiras, Vit\'oria, ES 29060-900, Brazil.}


%
\begin{abstract}
We investigate the existence of black bounce solutions in $2+1$ dimensions within the framework of $f(R)$ gravity. We analyze whether black bounce geometries originally obtained in GR can be consistently generalized to $f(R)$ theories and identify the matter sources capable of supporting such solutions. We also construct a new class of solutions by imposing a vanishing curvature scalar. In the matter sector, we consider models involving a coupling between a scalar field and nonlinear electrodynamics, while in the gravitational sector we analyze both the Starobinsky model and more general forms of $f(R)$. We further examine the viability conditions of the $f(R)$ models that give rise to these spacetimes, including the behavior of the scalaron mass. Finally, we study the associated energy conditions, in order to assess the degree of exoticity of the matter content required to sustain these black bounce solutions and how the $f(R)$ theory modifies the energy conditions.

\end{abstract}
\maketitle

\section{Introduction}
\label{intro}

There is broad consensus in the literature that the theory of general relativity (GR), proposed by Albert Einstein, admits a variety of models that exhibit issues related to geodesic singularities, thereby motivating the search for more general theories of gravity \cite{1,2,3,4,5,6,7,8,9,10}. From this perspective, conceptually, the singularities present in black holes (BHs) naturally emerge within the context of the of GR through the so-called gravitational collapse that occurs as a possibility for the final stage of a supermassive star. The simplest possible model proposed to describe a compact, spherically symmetric, and static object with such characteristics was presented by Karl Schwarzschild in 1916, as the geometric representation for a vacuum region \cite{11,12,13,14}. Subsequently, other solutions were proposed, such as the electrically charged Reissner-Nordström spacetime model representing the electrovacuum, models with a cosmological constant, and rotation scenarios such as the Kerr and Kerr-Newman geometries \cite{13,14}. From the perspective of models whose purpose is to remove singularities inside the BH, there are a variety of solutions for regular BHs that emerged around the 1970s. The first model was proposed by James Bardeen 1968 and the regulation parameter around the 2000s with Ayon-Beato and Garcia, gained an interpretation as being the charge of the magnetic monopole with a nonlinear electrodynamics (NED) \cite{10,15}. In addition to this model proposed by Bardeen, a wide variety of other regular geometries have been proposed over the years, with Rodrigues and Silva recently numerically developing Bardeen's version of electrically charged matter for spacetime \cite{16}. Beyond Bardeen's model, regular BHs solutions have been widely explored in a wide variety of scenarios \cite{17,18,19,20,21,22,23,24,25,26,27,28,29,30}.

In addition to the regular BH solutions discussed above, Simpson and Visser recently proposed a new class of regular geometries, known as black bounce (BB) \cite{Simpson:2018tsi}, constructed through a regularization procedure that removes geodesic singularities. The causal structure of these solutions depends strongly on the relationship between the free parameters of the model, in particular between the regularization parameter \textbf{a} and the mass \textbf{m}. For $a >2m$, the geometry describes a traversable wormhole (WH) with bidirectional passage. In the limiting case $a=2m$, one obtains a critical configuration corresponding to a unidirectional WH, whose throat is located at the origin of the coordinate system $r=0$. For $a<2m$, the solution represents a WH possessing two symmetrically positioned event horizons. In this context, this class of regular solutions has been widely investigated in several physical scenarios, including gravitational lensing, field dynamics, and stability analyses \cite{31,31A,32,33,34,35,36,37,38,39,40,41,42,43,44,45,45A,46,47,48,49,50}. Such analyses were expanded to modified gravity scenarios as per the $f(R)$ theories in (3+1) dimensions, where the approach was initiated by the authors in Ref.\cite{51} and subsequently an extended version was constructed \cite{52}. The cases investigated also included $f(T)$, $f(R,R_{\mu,\nu})$ \cite{53,54}, and $f(G)$ gravity \cite{54A,54B}, as well as k-essence theories \cite{55,56,57,58}. Still within the context of modified gravity, BB solutions were explored for systems where Cotton gravity is coupled to NED and scalar fields, as well as for conformal Killing gravity for the same couplings \cite{59,60}.

Regarding the investigation of BB solutions in (2+1) dimensions, the authors in \cite{61} analyzed the geometric scenarios starting from the BTZ BH model and then introduced the regularization process proposed by Simpson and Vissser \cite{Simpson:2018tsi}. In this work, the authors investigate both the usual case and the charged case, finding interesting results for geodesic stability for massive or non-massive particles. They showed that stable orbitals are only possible for the charged system, and an analysis of the energy conditions was also explored. Still in (2+1) dimensions, now in the context of modified gravity $f(R,R_{\mu,\nu})$, new BB solutions were analyzed for the asymptotically flat (zero cosmological constant) and asymptotically anti-de Sitter (AdS) cases. The authors investigated the matter source associated with the electrically charged system, where the need for the existence of NED, as well as a nonlinear scalar field, was verified. Finally, they analyzed the energy and stability conditions for the asymptotically flat case \cite{54}. Additionally, within the context of GR, the authors of Ref. \cite{62} analyze the appropriate scenario regarding the matter sources associated with electrically charged BH solutions, considering the regularized BTZ and Einstein–CIM geometries in (2+1) dimensions.

In Section \ref{EOM} of this work, we present the general relations and the derivation of the field equation for spherically symmetric spacetime in (2+1) dimensions. In Section \ref{21dimensions}, we analyze 4 types of $f(R)$ functions as well as the physical and geometric quantities for the electrically charged BB configuration obtained from the regularized BTZ BH. In Sections \ref{SEC:Viability} and \ref{EN}, the viability conditions, the mass of the scalaron, and the energy conditions are analyzed. Finally, in the \ref{conclusions} section, we present the final considerations and conclusions.

\section{Equations of motions in 2+1 dimensions}
\label{EOM}
We are interested in studying BBs in three dimensions within the framework of $f(R)$ gravity. It is common, at least in GR, that these objects have as their field sources a nonlinear electrodynamic minimally coupled to a scalar field. Therefore, we will consider the theory described by the action:
\begin{equation}\label{action}
    S = \int \sqrt{|g|}d^3x[f(R) - 2h(\phi)g^{\mu\nu}\partial_\mu\phi\partial_\nu\phi + 2V(\phi) + L(F)],
\end{equation}
where $f(R)$ is a function of the Ricci scalar $R = g^{\mu\nu}R_{\mu\nu}$, $\phi$ is the scalar field, $V(\phi)$ is potential associated to the scalar field, $L(F)$ is the NED Lagrangian, $F=F^{\mu\nu}F_{\mu\nu}$ is the electromagnetic scalar with $F^{\mu\nu}$ being the Maxwell-Faraday tensor, and $h(\phi)$ is a function that determines where the scalar field is phantom ($h(\phi) < 0$) or canonical ($h(\phi) > 0$).

To obtain the equations of motion, we must vary the action \eqref{action} with respect to $A_\mu$, $\phi$ and $g^{\mu\nu}$, and we get the following equations, respectively
\begin{equation}\label{Eq_Max}
    \nabla_\mu(L_F F^{\mu\nu}) = \frac{1}{\sqrt{|g|}}\partial_\mu(\sqrt{|g|}L_F F^{\mu\nu}) = 0,
\end{equation}
\begin{equation}\label{Eq_scalar}
    2h(\phi)\nabla_\mu\nabla^\mu\phi + \frac{dh(\phi)}{d\phi}\partial^\mu\phi\partial_\mu\phi = - \frac{dV(\phi)}{d\phi},
\end{equation}
\begin{equation}
    f_RR_{\mu\nu} - \frac{1}{2}g_{\mu\nu}f(R) + (g_{\mu\nu}\Box - \nabla_\mu\nabla_\nu)f_R = T[\phi]_{\mu\nu} + T[F]_{\mu\nu},\label{Eq_fR}
\end{equation}
where $L_F = dL/dF$, $f_R=df(R)/dR$, and $T[\phi]_{\mu\nu}$ and $T[F]_{\mu\nu}$ are the stress-energy tensors of the scalar and electromagnetic fields, given by
\begin{equation}
    T[F]_{\mu\nu} = \frac{1}{2}g_{\mu\nu}L(F) - 2L_FF^\alpha_\nu F_{\mu\alpha},
\end{equation}
\begin{equation}
    T[\phi]_{\mu\nu} = 2h(\phi)\partial_\mu\phi\partial_\nu\phi - g_{\mu\nu}(h(\phi)\partial^\alpha\phi\partial_\alpha\phi - V(\phi)).
\end{equation}

To begin solving the equations of motion, we must choose the form of our line element. We will choose the line element describing a static spacetime, written as:  
\begin{equation}
    ds^2=A(r)dt^2-\frac{1}{A(r)}dr^2-\Sigma^2(r)d\varphi^2.\label{ele}
\end{equation}
where the functions $A(r)$ and $\Sigma(r)$ are, in principle, arbitrary functions of the radial coordinate.  

It is possible to approach the problem in two different ways: we can either impose the source fields and the $f(R)$ gravity model to find the components of the metric tensor, or we can impose the components of the metric tensor and determine which functions for the source fields and the gravitational theory are most suitable to describe such solutions. Since modified gravitational theories like $f(R)$ involve fourth-order derivatives of the metric tensor components, it becomes more feasible to work with the second option.  

Since we are working with spacetimes in $2+1$ dimensions, it is more appropriate to consider that our object is electrically charged.  
Thus, considering that our solution has an electric charge and the line element \eqref{ele}, the only non-zero component of the Maxwell-Faraday tensor is $F^{10} =- F^{01}$. From the modified Maxwell equations \eqref{Eq_Max}, we find
\begin{equation}
      \frac{1}{\sqrt{\left|g\right|}}\frac{\partial }{\partial r}\left[\sqrt{\left|g\right|}L_F F^{10}\right]=0,\ 
  \quad \Rightarrow \quad \Sigma L_F F^{10}=q = constant,
\end{equation}
where $q$ is the electric charge. The electromagnetic scalar is written as
\begin{equation}
    F(r)=-\frac{2q^2}{\Sigma^2L_F^{2}}.
\end{equation}

With these considerations, the remaining equations of motion are given by:
\begin{eqnarray}
   && -2 h (\phi) A' \phi '-A \phi '(r) h'(\phi)-\frac{2 A h (\phi) \Sigma ' \phi '}{\Sigma
   }-2 A h (\phi) \phi ''=-\frac{V'}{\phi '},\label{eqmovphi}\\
   &&\frac{1}{2} f_R A''-\frac{1}{2} A' f_R'+\frac{f_R A' \Sigma '}{2 \Sigma
   }-A f_R''-\frac{A f_R' \Sigma '}{\Sigma }-A h (\phi) \phi
   '^2-\frac{f}{2}=\frac{L}{2}+\frac{2 q^2}{L_F \Sigma ^2}+V,\label{eqmov0}\\
   &&\frac{1}{2} f_R A''-\frac{1}{2} A' f_R'+\frac{f_R A' \Sigma '}{2 \Sigma
   }-\frac{A f_R' \Sigma '}{\Sigma }+\frac{A f_R \Sigma ''}{\Sigma }+A
   h (\phi) \phi '^2-\frac{f}{2}=\frac{L}{2}+\frac{2 q^2}{L_F \Sigma ^2}+V,\label{eqmov1}\\
   &&-A' f_R'+\frac{f_R A' \Sigma '}{\Sigma }-A f_R''+\frac{A f_R
   \Sigma ''}{\Sigma }-A h (\phi) \phi '^2-\frac{f}{2}=\frac{L}{2}+V.\label{eqmov2}
\end{eqnarray}
where $``\,'\,"$ represents the derivative with respect to the radial coordinate.

From equation \eqref{eqmov1} and \eqref{eqmov2}, we find the form of $L(r)$ and $L_F(r)$, which are given by
\begin{equation}\label{Lr}
   L(r) = -2 A' f_R'+\frac{2 f_R A' \Sigma '}{\Sigma }-2 A f_R''+\frac{2 A
   f_R \Sigma ''}{\Sigma }-2 A h (\phi) \phi '^2-f-2 V,
\end{equation}
\begin{equation}
    \label{Lfr}
    L_F(r) =-\frac{4 q^2}{\Sigma \left(-f_R \Sigma  A''-\Sigma  A' f_R'+f_R A' \Sigma '-2 A \Sigma f_R''+2 A f_R' \Sigma '-4 A \Sigma  h (\phi) \phi
   '^2\right)}.
\end{equation}

Using equations \eqref{Lr} and \eqref{Lfr}, the equation \eqref{eqmov0} becomes
\begin{equation}
      h (\phi) \phi '^2 = -\frac{\Sigma  f_R''+f_R \Sigma ''}{2\Sigma}.\label{Eq_hphi}
\end{equation}
Through this relation, we can infer the behavior of $h(\phi)\phi'^2$. As is customary in this type of situation, we will impose that our scalar field behaves as $\phi = \arctan(r/a)$, where $a$ is a constant that will be related with the regularization of the solutions, and we will determine the behavior of $h(\phi)$, which is written as
\begin{equation}
    h(\phi)=-\frac{\left(a^2+r^2\right)^2 \left(\Sigma  f_R''+f_R \Sigma ''\right)}{2 a^2 \Sigma }.
\end{equation}
With this relation, it is possible to deduce the behavior of $h(\phi)$ once we know the $f(R)$ theory we are dealing with and the $g_{22}$ coefficient of the metric.

An additional equation that we must take into account is the consistency relation, given by:
\begin{equation}\label{cons}
    L_F\frac{dF}{dr}-\frac{dL}{dr}=0.
\end{equation}
This equation acts as a constraint that the electromagnetic functions must satisfy.

Although the dynamics of the system is fundamentally described by the nonlinear Lagrangian $L(F)$, obtaining an explicit analytical form for $L(F)$ in the context of electrically charged BHs presents a significant challenge. While the electromagnetic invariant $F$ can be easily expressed as a function of the radial coordinate, $F(r)$, the complexity of most NED models prevents the algebraic inversion of this relation to obtain $r(F)$. Consequently, one cannot directly write the Lagrangian density in terms of the invariant $F$ in a closed analytical form. Furthermore, attempting such inversions often leads to multi-valued Lagrangians, where a single value of $F$ could correspond to different physical branches of the solution. To circumvent these difficulties, it is standard to adopt an alternative framework by introducing an auxiliary anti-symmetric tensor $P_{\mu\nu}$, defined as:
\begin{equation}P_{\mu\nu} \equiv L_F F_{\mu\nu}.
\end{equation}
We then define a structural function $H(P)$ through a Legendre transformation \cite{20}:\begin{equation}H(P) \equiv 2F L_F - L(F).\end{equation}In this dual representation, $P \equiv P^{\mu\nu}P_{\mu\nu}$ is the associated invariant. This approach effectively "unfolds" the multi-valued branches of the $L(F)$ formalism, providing a single-valued and well-behaved description of the theory in terms of $P$. From the definitions above, the following relations are established:
\begin{equation}P = (L_F)^2 F=-\frac{2q^2}{\Sigma^2}, \quad H_P L_F = 1, \quad \text{and} \quad F_{\mu\nu} = H_P P_{\mu\nu},
\end{equation}
where $H_P = dH/dP$. Using this dual formulation, the electromagnetic stress-energy tensor $T[F]^\mu_\nu$ can be written as:
\begin{equation}T[F]_{\mu\nu} = \frac{1}{2} g_{\mu\nu} \left( 2P H_P - H \right) - 2 H_P P_{\mu}^{\ \alpha} P_{\nu\alpha}.
\end{equation}
Finally, the original Lagrangian density can be recovered and analyzed as a function of the $P$ invariant:
\begin{equation}L(P) = 2P H_P - H.
\end{equation}
This $L(P)$ representation is particularly advantageous for constructing electrically charged BH solutions, as $P$ typically follows a simple, monotonic radial distribution, avoiding the inversion issues inherent to the $L(F)$ framework.

Thus, we have established all the necessary equations to solve our problem. In the next section, we will propose different approaches to address this problem.


\section{BBs in 2+1 dimensions}  \label{21dimensions}
As mentioned earlier, in general it is difficult to solve the field equations of $f(R)$ gravity to obtain a metric, since these equations involve fourth-order derivatives of the metric. However, we can exploit the functional freedom in the model to determine which sources would generate given metrics for specific $f(R)$ models.

For most of the approaches we will consider, it is quite feasible to use the regularized version of the BTZ metric, which is described by the line element \eqref{ele} with
\begin{equation}
    A(r)=-M+\frac{r^2+a^2}{l^2}, \qquad \mbox{and} \qquad \Sigma^2(r)=r^2+a^2,\label{metric_functions}
\end{equation}
where $l$ is related with the cosmological constant by $\Lambda=-1/l^2$. 
In the context of GR, this spacetime has been extensively studied, and we will adopt it as our model to carry out the generalizations in $f(R)$ gravity.

The curvature scalar related with this solution is written as
\begin{equation}
    R=\frac{4 a^4+10 a^2 r^2+6 r^4}{l^2 \left(a^2+r^2\right)^2}-\frac{2 a^2 M}{\left(a^2+r^2\right)^2}.\label{scalarR}
\end{equation}
This expression will be important later.

\subsection{Model \texorpdfstring{$f_R = 1 + a_2 r^2$}{}}
Intuitively, the first attempt to generalize a solution would be to propose a $f(R)$ theory model and check whether our spacetime is a solution. However, we will consider a different approach.

As in \cite{Silva:2025fqj}, we can consider different types of $f_R$ instead of working directly with the function $f(R)$. The function $f(R(r))$ will be obtained by
\begin{equation}
    f(R(r))=\int f_R\left(\frac{dR}{dr}\right)dr.
\end{equation}

In \cite{Rodrigues:2015ayd,Rodrigues:2016fym}, the authors show that, due to the symmetries of the field equations in $f(R)$ gravity and the form of the line element, the function $f_{R}=1+a_{1}r$ naturally arises from subtracting the components of the field equations. In our case, since the line element is more general and the field equations are more involved, this function does not arise naturally, which gives us a certain freedom in choosing it. An example of this type of proposal would be to consider the function $f_R$ as $f_R = 1 + a_n r^n$. However, geometric expressions become quite complicated, making it clearer to consider particular cases. At this moment, we will analyze the case $f_R = 1 + a_2 r^2$. One advantage of this form is that it is symmetric under the transformation $r \to -r$. The function $f(R(r))$ for this case is written as
\begin{equation}
    f(R(r))=-\frac{2 a^2 \left(a^2+l^2 M+r^2\right)}{l^2 \left(a^2+r^2\right)^2}+\frac{2 a_2}{l^2} \left(a^2 \ln \left(\frac{r^2}{a^2}+1\right)-\frac{r^2 \left(a^4+r^2 \left(a^2-l^2
   M\right)\right)}{\left(a^2+r^2\right)^2}\right),
\end{equation}
while the functions related to the field sources are
\begin{eqnarray}
    F(r)&=&-\frac{a^4 M^2}{2
   q^2 \left(a^2+r^2\right)^3}-\frac{a_2^2 M^2 r^4 \left(3 a^2+2 r^2\right)^2}{2 q^2 \left(a^2+r^2\right)^3}-\frac{a_2 M^2 a^2 r^2\left(3 a^2 +2  r^2\right)}{q^2 \left(a^2+r^2\right)^3},\\
    L(r)&=&\frac{4}{l^2}-\frac{3 a^2 M}{2 \left(a^2+r^2\right)^2}-2 a_2 M \ln \left(\frac{r^2}{a^2}+1\right)-\frac{a_2 M \left(8 a^2 r^2+5 r^4\right)}{2 \left(a^2+r^2\right)^2},\\
    L_F(r)&=&\frac{2 q^2 \left(a^2+r^2\right)}{M \left(a^2 \left(3 a_2 r^2+1\right)+2 a_2 r^4\right)},\\
        \phi(r)&=&\arctan\left(\frac{r}{a}\right),\quad
        h(r)=-\frac{1}{2}-a_2 \left(\frac{a^4+r^4}{a^2}+\frac{5 r^2}{2}\right),\\
    V(r)&=&\frac{a^2 \left(2 a^2+l^2 M+2 r^2\right)}{4 l^2 \left(a^2+r^2\right)^2}+\frac{a_2}{4 l^2 \left(a^2+r^2\right)^2} \left(-4 a^6+2 a^4 \left(2 l^2 M-9 r^2\right)\right.\nonumber\\
    &-&\left.4 \left(a^2+r^2\right)^2 \left(a^2-l^2 M\right) \ln
   \left(\frac{r^2}{a^2}+1\right)+2 a^2 \left(5 l^2 M r^2-13 r^4\right)+5 l^2 M r^4-12 r^6\right).
\end{eqnarray}

Expanding the function $h(r)$ for $r \to 0$, we obtain  
\begin{eqnarray}
    h(r)\approx -\frac{1}{2} \left(1+2 a^2 a_2\right) + O(r^2).
\end{eqnarray}
If the constant $a_2$ is positive, the scalar field will always be phantom. If $a_2$ takes negative values and if $-a^2a_2 > 1/2$, then we have a canonical scalar field. 

We can also invert the scalar field $\phi(r)$ and, consequently, express the functions $V(\phi)$ and $h(\phi)$, which are given by  
\begin{eqnarray}
    h(\phi)&=&-\frac{1}{2}+\frac{a^2 a_2}{4} \left(2 - (\cos 2 \phi +5) \sec ^4\phi \right),\\
    V(\phi)&=&\frac{\cos ^2\phi \left(4 a^2+l^2 M \cos 2 \phi +l^2 M\right)}{8 a^2 l^2}-\frac{a_2}{32 l^2} \left(4 \cos2 \phi \left(2 a^2+l^2 M\right)+32 a^2 \left(3 \sec ^2\phi +\ln
   \left(\sec ^2\phi \right)\right)\right.\nonumber\\
   &-&\left.72 a^2+l^2 M \left(\cos 4 \phi -32 \log \left(\sec ^2\phi \right)\right)-37 l^2 M\right).
\end{eqnarray}
The corrections from the $f(R)$ theory require much more nonlinearity in the matter sector compared to GR.

\begin{figure}
    \centering
    \includegraphics[width=0.5\linewidth]{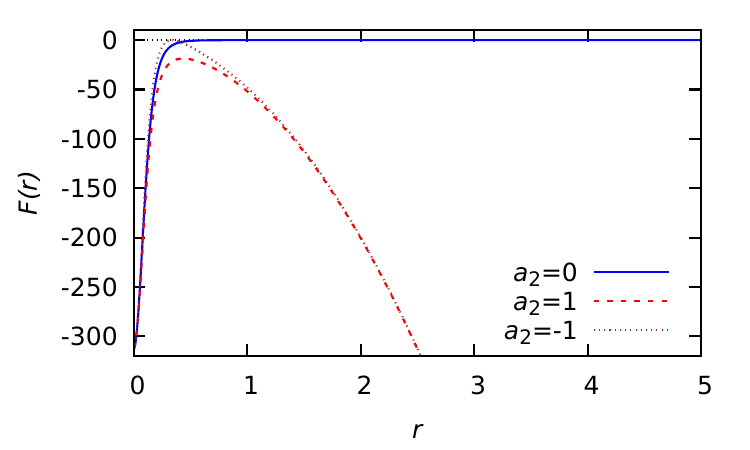}\hspace{-0.4cm}
    \includegraphics[width=0.5\linewidth]{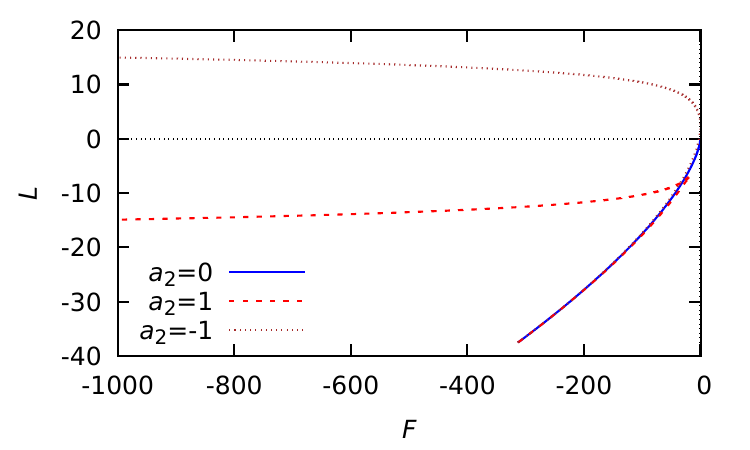}
    \caption{Behavior of the function $F(r)$ (left panel) and the electromagnetic Lagrangian $L(F)$ (right panel) for the parameters $M=1$, $q=a=0.2$, $l=10$, and different values of $a_{2}$ considering the model $f_R=1+a_2r^2$.}
    \label{fig:LF_model1}
\end{figure}

Although inverting $\phi(r)$ to express the functions in terms of the scalar field is straightforward, the situation is different in the electromagnetic sector. To clarify this point, we examine the behavior of $F(r)$ and $L(F)$ in Fig. \ref{fig:LF_model1}. As seen, for $a_{2}=0$ the function $F(r)$ has a minimum at $r=0$ and increases while tending to zero as $r$ grows. In principle, this case would allow us to invert and obtain $r(F)$, thereby deriving the behavior of $L(F)$ analytically. However, when the coupling term with $f(R)$ gravity is included, the behavior of $F(r)$ changes and $F(r)$ develops a maximum. Whenever $F(r)$ features maxima or minima, $L(F)$ becomes multivalued, precisely what we observe in Fig. \ref{fig:LF_model1}.
As mentioned previously, this type of issue arises when we adopt the $L(F)$ formalism to describe electrically charged solutions, since this is not the most appropriate framework. However, by considering the $H(P)$ formalism, we are able to obtain an explicit expression for the Lagrangian $L(P)$, which is given by
\begin{equation}
   L(P)= \frac{4}{l^2}-2 a_2 M \ln \left(-\frac{2 q^2}{a_2 P}\right)+\frac{M \left(3 a^4 a_2 P^2-a^2 P \left(4 a_2 q^2+3 P\right)-20 a_2 q^4\right)}{8 q^4}.
\end{equation}
In this way, we are able to express analytically the behavior of the electromagnetic Lagrangian and avoid the difficulties that arise when attempting to describe it in the $L(F)$ formalism.

Due to the simplicity of the model, it is possible to invert the function $R(r)$ and thereby obtain the function $f(R)$ associated with our problem, which is given by
\begin{equation}
f(R) =
-\frac{
6}{l^{2}} + R + 2a_2M
- \frac{2a_{2}\sqrt{a^{4}-2a^{2}l^{2}M\left(-6+l^{2}R\right)}
+ a^{2}a_{2}\left[
6-l^{2}R
+ 2\ln\!\left(
-\frac{a^{2}+\sqrt{a^{4}-2a^{2}l^{2}M\left(-6+l^{2}R\right)}}
{a^{2}\left(-6+l^{2}R\right)}
\right)
\right]
}{l^{2}} .
\end{equation}
Thus, our model features a linear term, corresponding to GR, a term associated with the cosmological constant (since the solution is not asymptotically flat), and, finally, the nonlinear terms of the theory.

\subsection{Model \texorpdfstring{$f_R = 1 + a_\Sigma \Sigma$}{}}
Another type of model we can consider is a function $f_R$ given by $f_R = 1 + a_\Sigma \Sigma^n$. This model is motivated by the regularization of Simpson-Visser, where $r \rightarrow \sqrt{r^2 + a^2}$, and, as in the previous case, it is symmetric under the transformation $r \rightarrow -r$. For simplicity, let's choose the case $n = 1$. In this way, the function $f(R(r))$ is given by
\begin{equation}
    f(R(r))=-\frac{2 a^2 \left(a^2+l^2 M+r^2\right)}{l^2 \left(a^2+r^2\right)^2}-\frac{4 a^2 a_\Sigma \left(3 a^2+2 l^2 M+3 r^2\right)}{3 l^2 \left(a^2+r^2\right)^{3/2}}.
\end{equation}

The functions related to the field sources are given by
\begin{eqnarray}
    F(r)&=&-\frac{a^4 M^2}{2 q^2 \left(a^2+r^2\right)^3}-\frac{a^2 a_\Sigma M^2}{q^2 \left(a^2+r^2\right)^{3/2}}-\frac{a_\Sigma^2 M^2}{2 q^2},\\
    L&=&\frac{4}{l^2}-\frac{3 a^2 M}{2 \left(a^2+r^2\right)^2},\quad
    L_F=\frac{2 q^2}{M \left(a_\Sigma \sqrt{a^2+r^2}+\frac{a^2}{a^2+r^2}\right)},\\
    \phi(r)&=&\arctan\left(\frac{r}{a}\right),\quad
    h(r)=-\frac{1}{2}-a_\Sigma\sqrt{a^2+r^2},\\
    V(r)&=&\frac{a^2 \left(2 a^2+l^2 M+2 r^2\right)}{4 l^2 \left(a^2+r^2\right)^2}+\frac{a^2 a_\Sigma \left(9 a^2+l^2 M+9 r^2\right)}{3 l^2 \left(a^2+r^2\right)^{3/2}}.
\end{eqnarray}
As in the previous cases, the presence of $f(R)$ theory requires more nonlinearities than in the case of GR. The analytical expressions for this case are relatively simpler than those in the previous cases. It is important to note that the constant $a_\Sigma$ does not appear in the function $L(r)$. What happens is that, for a general $n$, the contributions from $f(R)$ theory in the function $L(r)$ are proportional to $n-1$, so that, for this specific case, the contribution is zero. The functions $f(R)$, $V$, and $L$ exhibit divergences for $n = 2$ and $n = 4$. However, the combination $f(R) + L + 2V$, as it appears in the action of the theory, does not present these divergences. For positive $a_\Sigma$, the scalar field will always be phantom, whereas for negative $a_\Sigma$ and if $-a a_\Sigma > 1/2$, the scalar field will be canonical.

We can also write the functions $h(\phi)$ and $V(\phi)$ as
\begin{eqnarray}
    h(\phi)&=&-\frac{1}{2}-a_\Sigma a \sec ^2\phi,\\
    V(\phi)&=&\frac{1}{4} \cos ^2\phi  \left(\frac{M \cos ^2\phi }{a^2}+\frac{2}{l^2}\right)+\frac{a_\Sigma \cos\phi\left(9 a^2+l^2 M \cos ^2\phi \right)}{3 l^2 a }.
\end{eqnarray}

\begin{figure}
    \centering
    \includegraphics[width=0.5\linewidth]{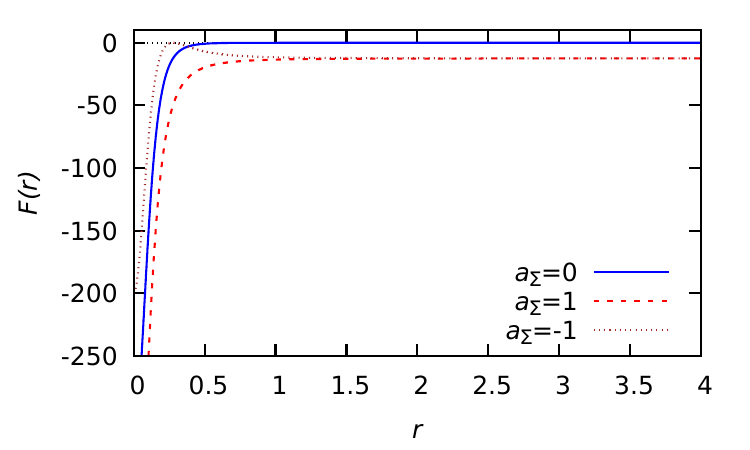}\hspace{-0.4cm}
    \includegraphics[width=0.5\linewidth]{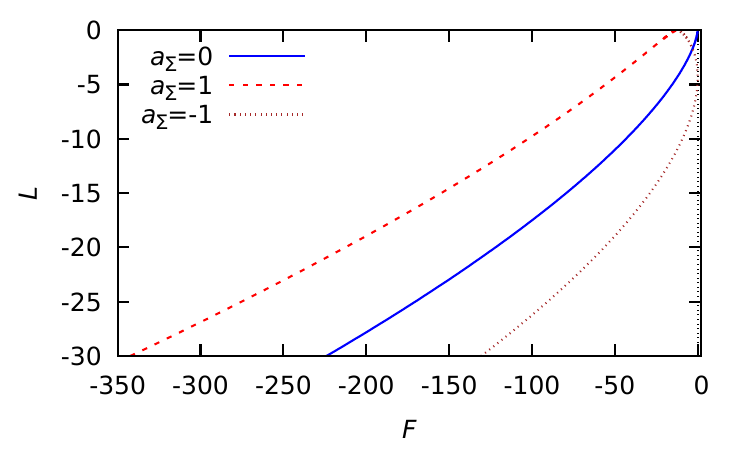}
    \caption{Behavior of the function $F(r)$ (left panel) and the electromagnetic Lagrangian $L(F)$ (right panel) for the parameters $M=1$, $q=a=0.2$, $l=10$, and different values of $a_{\Sigma}$, considering the model $f_R=1+a_\Sigma \Sigma$.}
    \label{fig:LF_model2}
\end{figure}
For the functions in the electromagnetic sector, we observe the same type of behavior as in the first case. In Fig. \ref{fig:LF_model2}, we show the behavior of $F(r)$ and the Lagrangian $L(F)$. We note that, for $a_{\Sigma}=0$, there are no maxima or minima and the function tends to zero at infinity; consequently, $L(F)$ is not multivalued. For $a_{\Sigma}=1$, $F(r)$ tends to a constant at infinity and likewise exhibits no extrema. However, for $a_{\Sigma}=-1$, $F(r)$ has a maximum, and thus we again observe a multivalued behavior in $L(F)$. Once again, in order to circumvent the multivalued behavior of the Lagrangian $L(F)$, we adopt the $H(P)$ formalism and express the electromagnetic Lagrangian in terms of $P$, namely $L(P)$, which is given by
\begin{equation}
    L(P)=\frac{4}{l^2}-\frac{3 a^2 M P^2}{2 q^4}.
\end{equation}
For the case $n=2$, we find that the Lagrangian $L(P)$ admits a relatively simple analytic form. However, it does not exhibit the Maxwell behavior, since it scales as $P^{2}$, whereas in Maxwell's theory one would obtain a linear dependence on $P$.

Thus, we managed to find all the functions related to the source fields for this $f_R$ function model.

Once again we can find the form of the $f(R)$ function of our theory, which is given by
\begin{equation}
  f(R)=  -\frac{6}{l^2}+R+\frac{4 a^2 a_\Sigma \left(l^2 R-6\right) \sqrt{\frac{a^2+\sqrt{a^4-2 a^2 l^2 M \left(l^2 R-6\right)}}{6-l^2 R}} \left(3
   a^2+3 \sqrt{a^4-2 a^2 l^2 M \left(l^2 R-6\right)}-2 l^4 M R+12 l^2 M\right)}{3 l^2 \left(a^2+\sqrt{a^4-2 a^2 l^2 M \left(l^2
   R-6\right)}\right)^2}.
\end{equation}
We have the term associated with the cosmological constant, the linear term corresponding to GR, and the nonlinear correction terms of the $f(R)$ theory.

\subsection{Model \texorpdfstring{$f(R) = R + a_R R^2$}{}}
We now consider a different approach. We will still use the regularized BTZ model, but instead of assuming a form for $f_{R}(r)$, we will adopt a known $f(R)$ gravity model. In this case, we consider the Starobinsky model, whose $f(R)$ function is given by:
\begin{equation}
    f(R)=R+a_R R^2,
\end{equation}
where $a_R$ is a constant. In the limit $a_R<<1$ we recover GR. 
The choice of the Starobinsky model is not accidental. This model stands out in the literature as one of the simplest and most successful extensions of GR \cite{Starobinsky:1980te}. Originally proposed to explain primordial inflation in a purely geometric way, it reproduces the observational data of the cosmic microwave background with remarkable accuracy. Moreover, the quadratic term in $R$ naturally arises as an effective quantum correction to the Einstein--Hilbert action, which provides the model with a strong theoretical motivation \cite{Birrell:1982ix}. In the context of astrophysical solutions, the Starobinsky model has been widely employed to investigate BHs, compact stars, and the effects of high-curvature corrections in $f(R)$ theories \cite{Olmo:2019flu,Sotiriou:2008rp}.

The function $f_R$, in this case, is
\begin{equation}
    f_R=\frac{df(R)}{dr}\left(\frac{dR}{dr}\right)^{-1}, \qquad \mbox{or} \qquad f_R=1+2a_R R.
\end{equation}
Thus, with \eqref{scalarR}, we have all necessary to determine the functions associated with the source fields, which are given by:
\begin{eqnarray}
    F(r)&=&-\frac{a^4 M^2}{2 q^2 \left(a^2+r^2\right)^3}-\frac{4 a^4 a_R M^2 \left(2 a^4+a^2 \left(7 r^2-l^2 M\right)+4 l^2 M r^2+5
   r^4\right)}{l^2 q^2 \left(a^2+r^2\right)^5}\nonumber\\
   &-&\frac{8 a^4 a_R^2 M^2 \left(2 a^4+a^2 \left(7 r^2-l^2 M\right)+4 l^2 M r^2+5 r^4\right)^2}{l^4 q^2
   \left(a^2+r^2\right)^7},\label{F_R2}\\
    L(r)&=&-\frac{2}{l^2}-\frac{3 a^2 M}{2 \left(a^2+r^2\right)^2}+a_R \left(\frac{12}{l^4}-\frac{30 a^2 M}{l^2 \left(a^2+r^2\right)^2}+\frac{20 a^2 M
   \left(3 a^2-4 l^2 M\right)}{3 l^2 \left(a^2+r^2\right)^3}+\frac{35 a^4
   M^2}{\left(a^2+r^2\right)^4}\right),\label{L_R2}\\
    L_F(r)&=&\frac{2 l^2 q^2 \left(a^2+r^2\right)^3}{a^2 M \left(l^2 \left(a^2+r^2\right)^2+4 a_R \left(2 a^4+a^2 \left(7
   r^2-l^2 M\right)+4 l^2 M r^2+5 r^4\right)\right)},\\
    \phi(r)&=&\arctan\left(\frac{r}{a}\right),\quad
     h(r)=-\frac{1}{2}-\frac{2 a_R \left(4 a^4+a^2 \left(3 l^2 M+r^2\right)-20 l^2 M r^2-3 r^4\right)}{l^2
   \left(a^2+r^2\right)^2},\label{hphi_model1}\\
    V(r)&=&\frac{a^2 \left(2 a^2+l^2 M+2 r^2\right)}{4 l^2 \left(a^2+r^2\right)^2}
    -a^2 a_R \left(\frac{3
    M }{ l^2 \left(a^2+r^2\right)^2}+\frac{2  M \left(9 a^2+40 l^2 M\right)}{3 l^2 \left(a^2+r^2\right)^3}+\frac{6
   }{l^4 \left(a^2+r^2\right)}-\frac{207 a^2 M^2}{6  \left(a^2+r^2\right)^4}\right).
\end{eqnarray}
Since we have chosen a specific form for the function $f(R)$, and consequently for $f_R(R)$, and the solution we are working with is not asymptotically flat, a constant term in the action is necessary. This is the reason why the constant term appears in $L(r)$.

If we expand $h(r)$, we find that
\begin{eqnarray}
    h(r)&\approx & -\frac{1}{2}-\frac{6 a_R M}{a^2}-\frac{8 a_R}{l^2}+O\left(r^2\right),\\
    h(r)&\approx& -\frac{1}{2}+\frac{6 a_R}{l^2}+O\left(\frac{1}{r^2}\right).
\end{eqnarray}
If we disregard the corrections from $f(R)$, i.e., the case of GR, the scalar field will always be phantom. When corrections from the $f(R)$ theory are considered, the sign and value of $a_R$ will be fundamental in determining the regions where the scalar field will be phantom. If $a_R > 0$, the scalar field will be phantom in more central regions and exhibit canonical behavior at infinity if $12a_R > l^2$. If $a_R < 0$, the scalar field can exhibit canonical behavior near $r = 0$ if $-a_R > a^2 l^2/(16a^2 + 12l^2 M)$ and will be phantom at infinity.

It is possible to write  $h(\phi)$ and $V(\phi)$ analytically, which are given by:
\begin{eqnarray}
 h(\phi)&=&-\frac{1}{2}+a_R \left(-\cos 2 \phi  \left(\frac{3 M}{a^2}+\frac{7}{l^2}\right)-\frac{23 M \cos 4 \phi }{4 a^2}+\frac{11
   M}{4 a^2}-\frac{1}{l^2}\right),\\
    V(\phi)&=&\frac{1}{4} \cos ^2\phi \left(\frac{M \cos ^2\phi}{a^2}+\frac{2}{l^2}\right)\nonumber\\
    &+&\frac{a_R \cos ^2\phi}{6 l^4} 
   \left(\frac{l^2 M \cos ^2\phi \left(-4 \cos ^2\phi \left(9 a^2+40 l^2 M\right)-18 a^2+207 l^2 M \cos ^4\phi
   \right)}{a^4}-36\right).\nonumber\\
\end{eqnarray}

\begin{figure}
    \centering
    \includegraphics[width=0.5\linewidth]{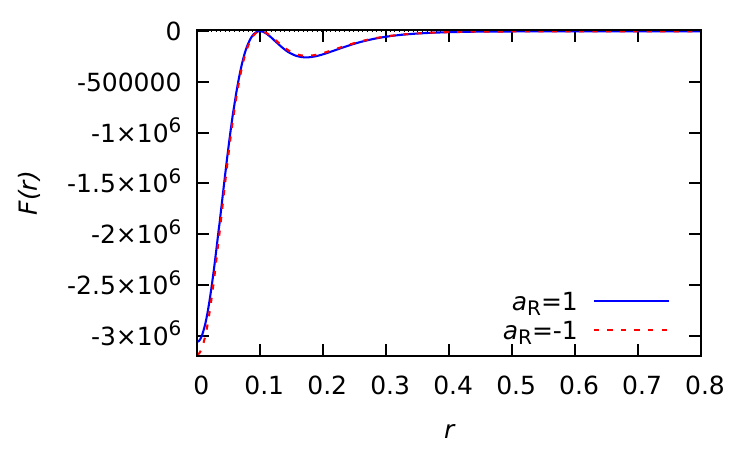}\hspace{-0.4cm}
    \includegraphics[width=0.5\linewidth]{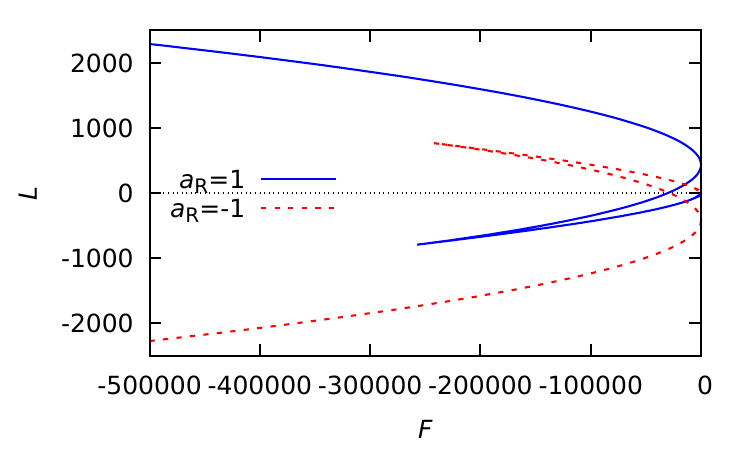}
    \caption{Behavior of the function $F(r)$ (left panel) and the electromagnetic Lagrangian $L(F)$ (right panel) for the parameters $M=1$, $q=a=0.2$, $l=10$, and different values of $a_{R}$, considering the Starobinsky gravity.}
    \label{fig:LF_model3}
\end{figure}

As seen from Eqs.~\eqref{F_R2} and \eqref{L_R2}, the expressions in the electromagnetic sector are even more involved than in the previous cases. In Fig.~\ref{fig:LF_model3}, the function $F(r)$ now exhibits one maximum and one minimum; consequently, three branches of $F(r)$ arise, separated by these two points. These three branches lead to three distinct behaviors in the Lagrangian $L(F)$, making the multivalued character even more evident than in the previous cases. Once again, we circumvent the issue of multivalued functions by adopting the $H(P)$ formalism, which is more suitable for electrically charged configurations. In this way, we express the electromagnetic Lagrangian in terms of $P$, namely $L(P)$, which is given by
\begin{equation}
    L(P)=\frac{12 a_R}{l^4}-\frac{2}{l^2}.+\frac{35 a^4 a_R M^2 P^4}{q^8}+\frac{20 a^2 a_R M P^3 \left(4 l^2 M-3 a^2\right)}{3 l^2 q^6}-\frac{3 a^2 M P^2 \left(20
  a_R+l^2\right)}{2 l^2 q^4}.
\end{equation}
Although this model is more complex than the previous cases, it is still possible to obtain the Lagrangian $L(P)$ in analytic form. It is worth noting that most of the nonlinear terms arise precisely from the coupling with the $f(R)$ theory, namely those terms proportional to the constant $a_{R}$.

Thus, we demonstrate that the regularized BTZ solution can be obtained through the $f(R)$ theory if we consider the coupling with a scalar field along with a NED. It becomes evident that the presence of the nonlinear term in the gravitational theory further necessitates complexities in the matter sector.

\subsection{Model using \textit{R=0}}
We now adopt a different approach as the final case. We still consider the Starobinsky model, but we impose that our solution has vanishing curvature scalar, i.e., $R=0$. Accordingly, in Starobinsky gravity, the function $f(R)$ and its derivative are given by
\begin{equation}
    f(R)=0, \quad \quad f_R(R)=1, \mbox{and} \quad f_{RR}(R)=2a_R.
\end{equation}
Considering again that the area function is given by $\Sigma^2=r^2+a^2$ and using all the equations of motion with the equation of consistence \eqref{cons}, we find that $A(r)$ is given by
\begin{equation}
    A(r)=c_1 \cos \left(\sqrt{2} \tan ^{-1}\left(\frac{r}{a}\right)\right)+c_2 \sin \left(\sqrt{2} \tan ^{-1}\left(\frac{r}{a}\right)\right),
\end{equation}
where $c_1$ and $c_2$ are constants.

We can attempt to determine the constants by imposing asymptotic limits on the solution. However, neither the BTZ solution nor its charged version could be used. An alternative is to consider the cosmological constant free version of the Einstein-CIM solution, given by
\begin{equation}
    A(r)=-M-\frac{Q}{2r}, \quad \Sigma=r,\quad Q:=(2q^2)^{3/4},
\end{equation}
where $M$ plays the role of an effective mass, and $q$ is the electric charge.
Thus, assuming that our solution behaves as Einstein-CIM at large distances, we obtain
\begin{equation}
    A(r)=- M \cos \left(\frac{\pi -2 \tan ^{-1}\left(\frac{r}{a}\right)}{\sqrt{2}}\right)-\frac{\sqrt{2} Q \sin \left(\frac{\pi -2 \tan
   ^{-1}\left(\frac{r}{a}\right)}{\sqrt{2}}\right)}{4 a}.\label{Amodel4}
\end{equation}
We can also expand this expression for $r \to 0$ and obtain\begin{equation}
    A(r)\approx -\frac{4 a M \cos \left(\frac{\pi }{\sqrt{2}}\right)+\sqrt{2} Q \sin \left(\frac{\pi }{\sqrt{2}}\right)}{4 a}+\frac{r \left(Q \cos
   \left(\frac{\pi }{\sqrt{2}}\right)-2 \sqrt{2} a M \sin \left(\frac{\pi }{\sqrt{2}}\right)\right)}{2 a^2}+O\left(r^2\right).
\end{equation}
This shows that the metric coefficients are regular at $r=0$.

\begin{figure}
    \centering
    \includegraphics[width=0.5\linewidth]{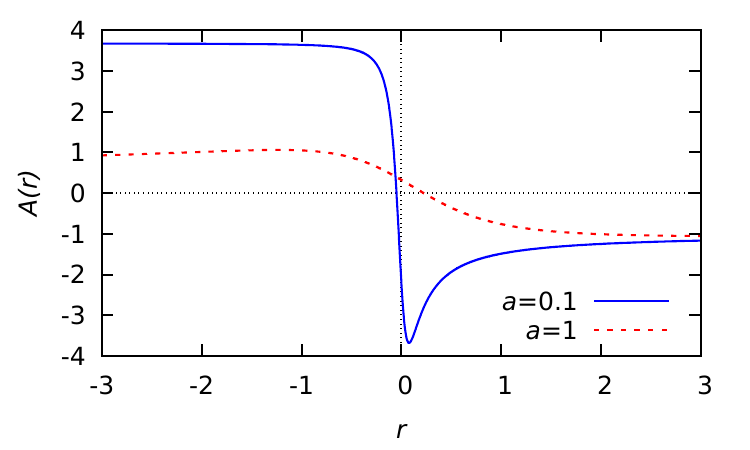}
    \caption{Behavior of the function $A(r)$, Eq. \eqref{Amodel4}, in terms of the radial coordinate with $M=Q=1$.}
    \label{fig:Amodel4}
\end{figure}

In Fig.~\ref{fig:Amodel4},  we illustrate the behavior of the metric function $A(r)$. For the chosen signature $(+, -, -, -)$, the static region of the spacetime is located where $A(r) > 0$. In this domain, the temporal Killing vector $\partial_t$ remains timelike, ensuring that the radial coordinate $r$ represents a spatial distance. As $r$ increases towards the horizon $r_h$, $A(r)$ vanishes and subsequently becomes negative for $r > r_h$. This transition marks the 'inverted' nature of the solution: unlike the standard Schwarzschild case where the exterior is at $r \to \infty$, here the region $r \to \infty$ corresponds to a non-static interior where the coordinates $t$ and $r$ exchange their causal roles. Therefore, the physically accessible universe for a stationary observer is restricted to the region $r < r_h$, while the domain $r > r_h$ describes the BH interior. This type of solution has been studied previously in the context of cylindrical BHs \cite{Bronnikov:2023aya,Bronnikov:2019clf,Bronnikov:1979ldq}.

To assess the regularity of the spacetime, we can analyze the Kretschmann scalar of this solution. The analytic expression is somewhat involved, but we can consider the following limits:
\begin{equation}
\begin{split}
    K&\approx \frac{\left(4 a M \cos \left(\frac{\pi }{\sqrt{2}}\right)+\sqrt{2} Q \sin \left(\frac{\pi }{\sqrt{2}}\right)\right)^2}{2 a^6}+O\left(r\right),\quad \mbox{if} \quad r\to 0,\\
   K &\approx \frac{3 Q^2}{2 r^6}-\frac{18 \left(a^2 M Q\right)}{r^7}+O\left(\frac{1}{r^8}\right),\quad \mbox{if} \quad r\to\infty.
   \end{split}
\end{equation}
Thus, we conclude that there are no curvature singularities in this spacetime.

The functions related to the field sources are given by
\begin{eqnarray}
    F(r)&=&-\frac{\left(\left(8 a^2 M-3 Q r\right) \cos \left(\frac{\pi -2 \tan ^{-1}\left(\frac{r}{a}\right)}{\sqrt{2}}\right)+2 \sqrt{2} a (3 M r+Q) \sin
   \left(\frac{\pi -2 \tan ^{-1}\left(\frac{r}{a}\right)}{\sqrt{2}}\right)\right)^2}{32 q^2 \left(a^2+r^2\right)^3},\label{F_model4}\\
    L(r)&=&-\frac{1}{28 a^3 \left(a^2+r^2\right)^2}\left[2 a \left(46 a^4 M+2 a^2 r (9 M r-13 Q)+9 r^3 (2 M r-Q)\right) \cos \left(\frac{\pi -2 \tan
   ^{-1}\left(\frac{r}{a}\right)}{\sqrt{2}}\right)\right.\nonumber\\
   &+&\left.\sqrt{2} \left(a^4 (104 M r+23 Q)+9 a^2 r^2 (4 M r+Q)+9 Q r^4\right) \sin \left(\frac{\pi -2
   \tan ^{-1}\left(\frac{r}{a}\right)}{\sqrt{2}}\right)\right],\label{L_model4}\\
    L_F(r)&=&\frac{8 q^2 \left(a^2+r^2\right)}{\left(8 a^2 M-3 Q r\right) \cos \left(\frac{\pi -2 \tan ^{-1}\left(\frac{r}{a}\right)}{\sqrt{2}}\right)+2
   \sqrt{2} a (3 M r+Q) \sin \left(\frac{\pi -2 \tan ^{-1}\left(\frac{r}{a}\right)}{\sqrt{2}}\right)},\\
    \phi(r)&=&\arctan\left(\frac{r}{a}\right),\quad
     h(r)=-\frac{1}{2},\label{hphi_model4}\\
    V(r)&=&\frac{\left(4 a^4 M+6 a^2 r (3 M r-2 Q)+9 r^3 (2 M r-Q)\right) \cos \left(\frac{\pi -2 \tan
   ^{-1}\left(\frac{r}{a}\right)}{\sqrt{2}}\right)}{28 a^2 \left(a^2+r^2\right)^2}\nonumber\\
   &+&\frac{\left(2 a^4 (24 M r+Q)+9 a^2 r^2 (4 M r+Q)+9 Q r^4\right) \sin \left(\frac{\pi -2 \tan
   ^{-1}\left(\frac{r}{a}\right)}{\sqrt{2}}\right)}{28 \sqrt{2} a^3 \left(a^2+r^2\right)^2}.
\end{eqnarray}
As we can see, this solution features a scalar field that is always phantom. 

\begin{figure}
    \centering
    \includegraphics[width=0.5\linewidth]{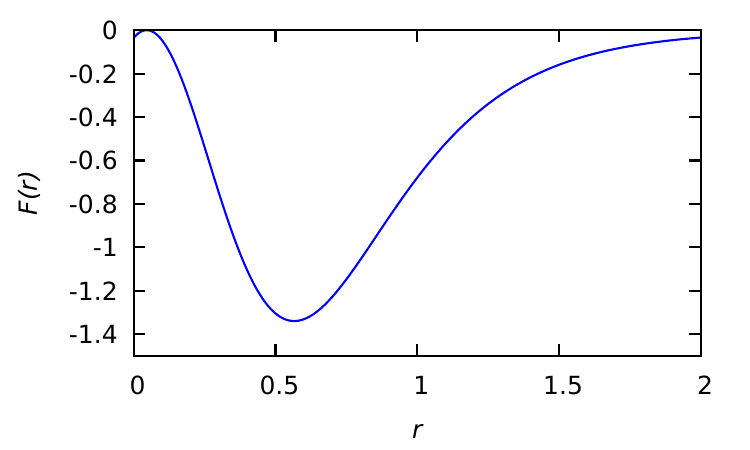}\hspace{-0.4cm}
    \includegraphics[width=0.5\linewidth]{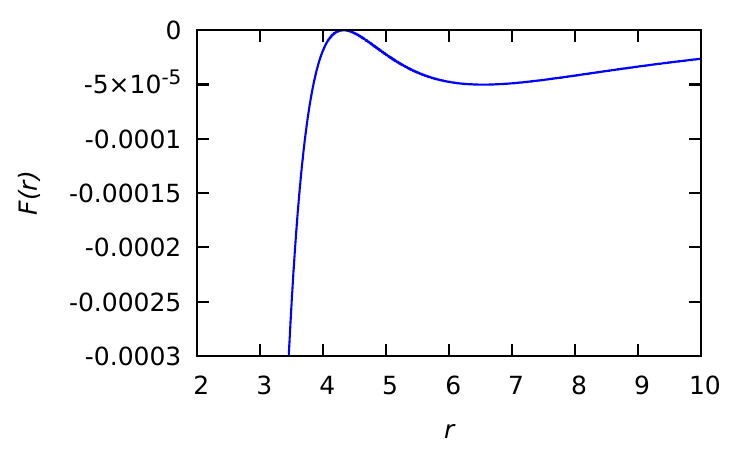}\hspace{-0.4cm}
    \includegraphics[width=0.5\linewidth]{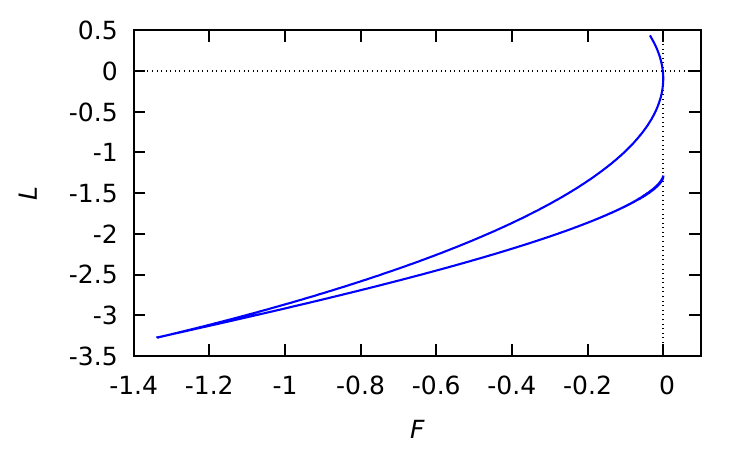}\hspace{-0.4cm}
    \includegraphics[width=0.5\linewidth]{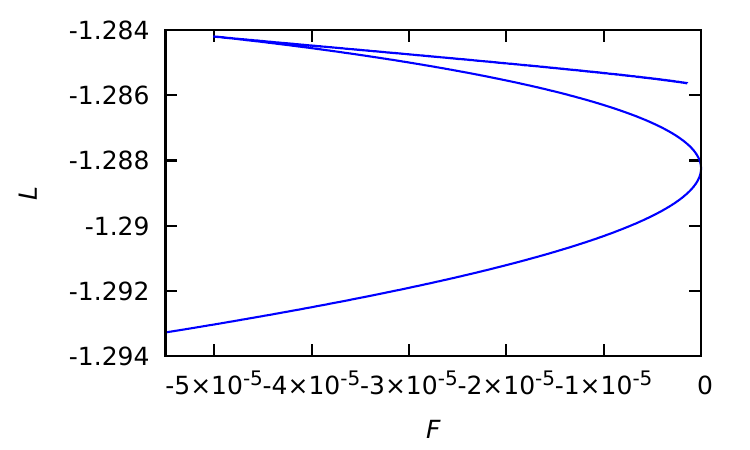}
    \caption{Behavior of the function $F(r)$ and the electromagnetic Lagrangian $L(F)$ for the parameters $M=1=q=a=1$, considering the Starobinsky gravity and $R=0$.}
    \label{fig:LF_model4}
\end{figure}

We can still invert the scalar field $\phi(r)$ and thereby obtain an analytic expression for the potential $V(\phi)$, which is given by
\begin{align}
\alpha(\phi) &\equiv \frac{\pi-2\phi}{\sqrt{2}},\\
V(\phi) &= \frac{1}{112\,a^{3}}\Big[
\big(\cos^{2}(2\phi)-7\cos(2\phi)+10\big)\Big(4aM\cos\alpha+\sqrt{2}\,Q\sin\alpha\Big)\nonumber\\
&\hspace{2.6cm}
+\sin(2\phi)\big(7+\cos(2\phi)\big)\Big(-3Q\cos\alpha+6\sqrt{2}\,aM\sin\alpha\Big)
\Big].
\end{align}
As for the electromagnetic sector, it is not possible to write an analytic form for $L(F)$. Nevertheless, we can discuss it further using Fig.~\ref{fig:LF_model4}. We see that the electromagnetic scalar $F(r)$ has several extrema, and each extremum implies a change in the behavior of the Lagrangian $L(F)$, as can also be seen in the figure. Thus, both in the behavior of $V(\phi)$ and of $L(F)$, we see that the matter sector requires several nonlinearities to sustain this BB solution.
The nonlinearity of the electromagnetic sector can also be expressed through the Lagrangian $L(P)$, which is given by
\begin{equation}
    L(P) = \frac{1}{14 a^3 q^4} \Big[ \mathcal{C} \cos(\theta) + \mathcal{S} \sin(\theta) \Big],
\end{equation}
where
\begin{equation}
\begin{split}
   & \theta = \frac{1}{\sqrt{2}} \left[ \pi + 2i \operatorname{arctanh} \left( \frac{\Delta}{a\sqrt{-P}} \right) \right], \quad\Delta = \sqrt{-a^2 P - q^2}, \quad \Phi = i \sqrt{-P} \Delta, \\
    &\mathcal{C} = -a \left[ 2M(23a^4 P^2 + 9a^2 P q^2 + 9q^4) + 9 \cdot 2^{3/4} (q^2)^{3/4} \Phi \left( q^2 - \frac{17}{9} a^2 P \right) \right], \\
     &\mathcal{S} = 2^{1/4} \left[ -9 (q^2)^{3/4} (q^4 + a^2 P q^2) - 23 a^4 P^2 (q^2)^{3/4} - 34 \sqrt{2} M a^4 P \Phi + 18 \sqrt{2} M a^2 q^2 \Phi \right].
\end{split}
\end{equation}
In this way, we see that, from both the $L(F)$ and the $H(P)$ formalisms, this model requires a significantly higher degree of nonlinearity in the electromagnetic sector than the previous models.

Thus, we have shown that a combination of NED with a phantom scalar field can, within $f(R)$ gravity, yield a spacetime representing an inverted BH whose curvature scalar vanishes.

Through these results, we may now perform a more detailed analysis and provide a classification of this new solution. We have seen that no singularity exists throughout the entire spacetime. We have also found that there is only a single horizon, whose location depends on the choice of the parameters, namely the charge and the mass. Considering our Universe to lie in the positive branch of the radial coordinate, see Fig.~\ref{fig:Amodel4} (red curve), we may have a horizon located at positive $r$, that is, $r_H>0$, together with a throat at $r=0$\footnote{A throat is defined as a non-vanishing minimum of the area function $\Sigma^2(r)$, with metric signature $(+--)$.}. Likewise, one may have a horizon in the negative branch of $r$, that is, $r_H<0$, see Fig.~\ref{fig:Amodel4} (blue curve), and a bounce at $r=0$\footnote{A bounce is defined as a non-vanishing minimum of the area function $\Sigma^2(r)$, with metric signature $(-+-)$.}. Therefore, since no singularity is present and there exists a horizon together with a minimum of $\Sigma^2(r)$, while $A(r)$ is not symmetric in $r$, the solution may be classified as follows: (a) an inverted asymmetric BB\footnote{See the general definitions for spherically symmetric metrics in Section IV of Ref.~\cite{rois2025novel}.} if our Universe lies in the positive branch of $r$; (b) an asymmetric BB if our Universe lies in the negative branch of $r$. In classification (a), one may have the throat located before the event horizon, or a bounce occurring after the event horizon. In classification (b), the situation is analogous, but reversed.

\section{Viability condition and scalaron mass}\label{SEC:Viability}

For an $f(R)$ model to be phenomenologically viable, it must satisfy a set of standard consistency requirements. The conditions are
$f_R(R)\equiv df(R)/dR>0$, ensuring a positive effective gravitational coupling (and excluding ghost-like tensor modes), and
$f_{RR}(R)\equiv d^2f(R)/dR^2>0$, which avoids the Dolgov--Kawasaki instability. 

We begin with the $D$-dimensional action
\begin{equation}
S=\frac{1}{2\kappa^2}\int d^D x \sqrt{-g}\, f(R)
+ S_{\text{mat}}.
\end{equation}
Variation with respect to the metric yields the field equations
\begin{equation}
f_R R_{\mu\nu}
-\frac12 g_{\mu\nu} f
+\left(g_{\mu\nu}\Box-\nabla_\mu\nabla_\nu\right) f_R
=\kappa^2 T_{\mu\nu}.
\end{equation}
Taking the trace and specialising to $D=3$, one obtains
\begin{equation}
f_R R
-\frac{3}{2} f
+2\Box f_R
=\kappa^2 T.
\end{equation}
We define the trace functional
\begin{equation}
E =
f_R R
-\frac{3}{2} f
+2\Box f_R
-\kappa^2 T.
\end{equation}
The dynamics is therefore encoded in
\begin{equation}
E=0.
\end{equation}
Performing a functional variation around a background configuration gives
\begin{equation}
\delta E=0.
\end{equation}
The relevant variations are
\begin{equation}
\delta f = f_R \delta R,
\delta f_R = f_{RR}\delta R,
\delta(\Box f_R)=\Box(f_{RR}\delta R).
\end{equation}
It follows that
\begin{equation}
\delta(f_R R) = f_{RR}R\,\delta R + f_R\delta R,
\delta\!\left(-\frac32 f\right) = -\frac32 f_R\delta R,
\delta(2\Box f_R) = 2\Box(f_{RR}\delta R).
\end{equation}
Collecting terms, the linearised trace equation reads
\begin{equation}
\left[
f_{RR}R
-\frac12 f_R
\right]\delta R
+2\Box(f_{RR}\delta R)
=\kappa^2 \delta T.
\end{equation}
We now expand around a generally non-constant background curvature,
\begin{equation}
R(r)=R_0(r)+\delta R.
\end{equation}
The d'Alembertian term expands as
\begin{equation}
\Box(f_{RR}\delta R)
=
f_{RR}\Box\delta R
+2\nabla^\mu f_{RR}\nabla_\mu \delta R
+(\Box f_{RR})\delta R.
\end{equation}
All functions of $R$ are evaluated on $R_0(r)$. The linearised equation therefore becomes
\begin{equation}
2 f_{RR}\Box\delta R
+4\nabla^\mu f_{RR}\nabla_\mu \delta R
+
\left[
f_{RR}R_0
-\frac12 f_R
+2\Box f_{RR}
\right]\delta R
=
\kappa^2 \delta T.
\end{equation}
Introducing the scalaron field
\begin{equation}
\Psi\equiv\delta R,
\end{equation}
we obtain the full differential equation governing the scalar degree of freedom,
\begin{equation}
2 f_{RR}\Box\Psi
+4\nabla^\mu f_{RR}\nabla_\mu \Psi
+
\left[
f_{RR}R_0
-\frac12 f_R
+2\Box f_{RR}
\right]\Psi
=
\kappa^2 \delta T.
\end{equation}
Dividing by $2 f_{RR}$ leads to
\begin{equation}
\Box\Psi
+
2\frac{\nabla^\mu f_{RR}}{f_{RR}}\nabla_\mu \Psi
-
m_{\Psi}^2(r)\Psi
=
\frac{\kappa^2}{2f_{RR}}\delta T.
\end{equation}
The effective scalaron mass squared is therefore (where $R_0(r)\rightarrow R(r)$, with $R(r)$  being the Ricci scalar of the background spacetime.)\footnote{In $4D$, the expression for the scalaron mass can be found in the well-known review \cite{Sotiriou:2008rp}, see equation (97) and the references therein.}
\begin{equation}
m_{\Psi}^2(r)
=
\frac{
\frac12 f_R
- f_{RR}R
-2\Box f_{RR}
}{
2 f_{RR}
}.\label{ms}
\end{equation}
Hence, in three-dimensional $f(R)$ gravity with a non-constant background, the scalaron obeys a modified Klein–Gordon-type equation with derivative couplings and a spatially dependent effective mass entirely determined by the function $f(R)$ and the background curvature. The sign of $m_{\psi}^{2}$ controls the local curvature of the scalaron effective potential around the background: $m_{\Psi}^{2}>0$ corresponds to a local minimum and signals stability against scalar perturbations, whereas $m_{\Psi}^{2}<0$ indicates tachyonic behavior.

\subsection{Model \texorpdfstring{$f_R = 1 + a_2 r^2$}{}}
For case $f_R = 1 + a_2 r^2$, the first viability condition, $f_R>0$, is trivially satisfied provided that $a_2>0$.
To verify the second condition, $f_{RR}>0$, we first need to obtain $f_{RR}$, which is given by
\begin{equation}
    f_{RR}=\frac{a_2 l^2 \left(a^2+r^2\right)^3}{2 a^2 \left(a^2+2 l^2 M+r^2\right)}.
\end{equation}
For $M>0$ and $a_{2}>0$, condition $f_{RR}>0$ is always satisfied.

In relation to the mass of the scalaron, we find
\begin{eqnarray}
m_{\Psi}^{2}=&&\frac{1}{2} \Bigg[\frac{a^2 \left(a^2 (-a_2)+64 a_2l^2 M+1\right)}{a_2 l^2
   \left(a^2+r^2\right)^2}-\frac{2 a^2 M \left(a^2 a_2-1\right)}{a_2 \left(a^2+r^2\right)^3}+\frac{-107
   a^2-72 l^2 M}{l^2 \left(a^2+r^2\right)}+\frac{2 \left(35 a^2+52 l^2 M\right)}{l^2 \left(a^2+2 l^2
   M+r^2\right)}\nonumber\\
&&+\frac{48 M \left(a^2+2 l^2 M\right)}{\left(a^2+2 l^2 M+r^2\right)^2}+\frac{42}{l^2}\Bigg].
\end{eqnarray}
Expanding to $r\to\infty$, we find that
\begin{equation}    m_{\Psi}^2\approx\frac{21}{l^2}+\frac{32 l^2 M-37 a^2}{2 l^2 r^2}+O\left(r^{-3}\right).
\end{equation}
We see that the scalaron mass is positive at radial infinity. We can always render the effective scalaron mass positive throughout the entire spacetime by appropriately adjusting the parameter values. We illustrate one such case in Fig.~\ref{msmods}~a), where we choose $M = 2$, $a = 1$, $l = 2$, and $a_{2}=0.005$. In this case, the model is free from tachyonic instability.
\subsection{Model \texorpdfstring{$f_R = 1 + a_\Sigma \Sigma$}{}}
For the case $f_R = 1 + a_\Sigma \Sigma$, the first viability condition, $f_R>0$, is trivially satisfied provided that $a_\Sigma>0$.

We find that the function $f_{RR}$ is given by
\begin{equation}
    f_{RR}=\frac{a_\Sigma l^2 \left(a^2+r^2\right)^{5/2}}{4 a^2 \left(a^2+2 l^2 M+r^2\right)}.
\end{equation}
This implies that the condition $f_{RR}>0$ is satisfied when $M>0$ and $a_\Sigma>0$.

Checking now the scalaron mass, we obtain that
\begin{eqnarray}
m_{\Psi}^2=&&\frac{a^2}{\text{a$_\Sigma $} l^2 \left(a^2+r^2\right)^{3/2}}+\frac{2 a^2 M}{\text{a$_\Sigma $}
   \left(a^2+r^2\right)^{5/2}}+\frac{6 a^2 M+64 M r^2}{\left(a^2+r^2\right) \left(a^2+2 l^2 M+r^2\right)}-\frac{24 M
   r^2}{\left(a^2+2 l^2 M+r^2\right)^2}\nonumber\\
   &&+\frac{2 \left(a^2+6 r^2\right)}{l^2 \left(a^2+r^2\right)}-\frac{M \left(2
   a^2+25 r^2\right)}{\left(a^2+r^2\right)^2}.
\end{eqnarray}
Expanding to $r\to\infty$, we find that
\begin{equation}
    m_{\Psi}^2\approx \frac{12}{l^2}+\frac{15 M-\frac{10 a^2}{l^2}}{r^2}+O\left(r^{-3}\right).
\end{equation}
Here again, we can always choose the parameters such that the effective scalaron mass remains positive throughout the entire spacetime. We illustrate the case $M = 2$, $a = 1$, $l = 2$, and $a_{\Sigma}=0.03$ in Fig.~\ref{msmods}~b).
\subsection{Model \texorpdfstring{$f(R) = R + a_R R^2$}{}}
\begin{figure}[htbp]
\centering

\begin{subfigure}{0.48\columnwidth}
\centering
\includegraphics[width=\linewidth]{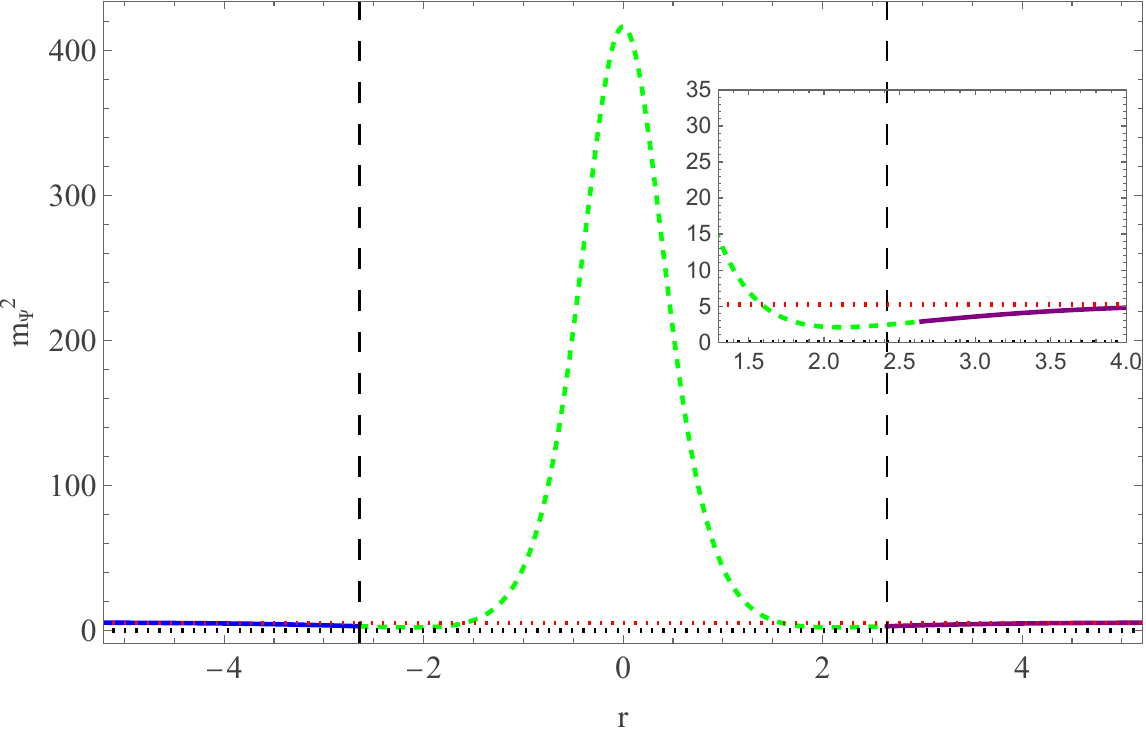}
\caption{}
\end{subfigure}
\hfill
\begin{subfigure}{0.48\columnwidth}
\centering
\includegraphics[width=\linewidth]{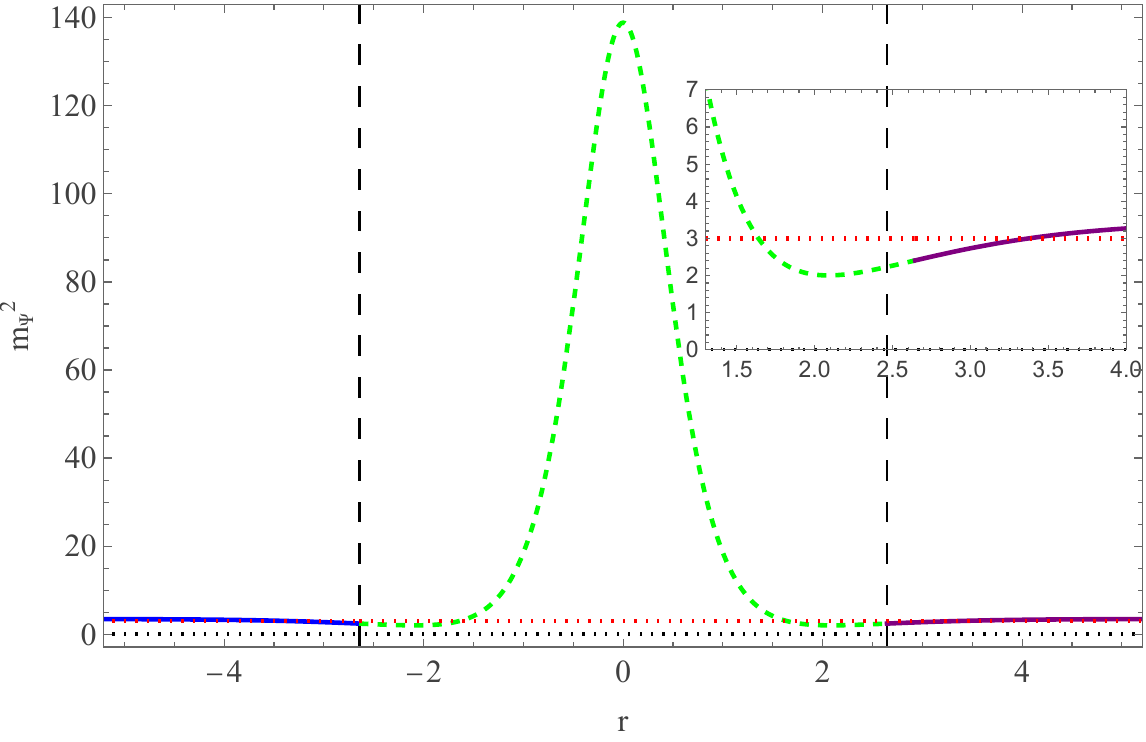}
\caption{}
\end{subfigure}

\vspace{0.3cm}

\begin{subfigure}{0.6\columnwidth}
\centering
\includegraphics[width=\linewidth]{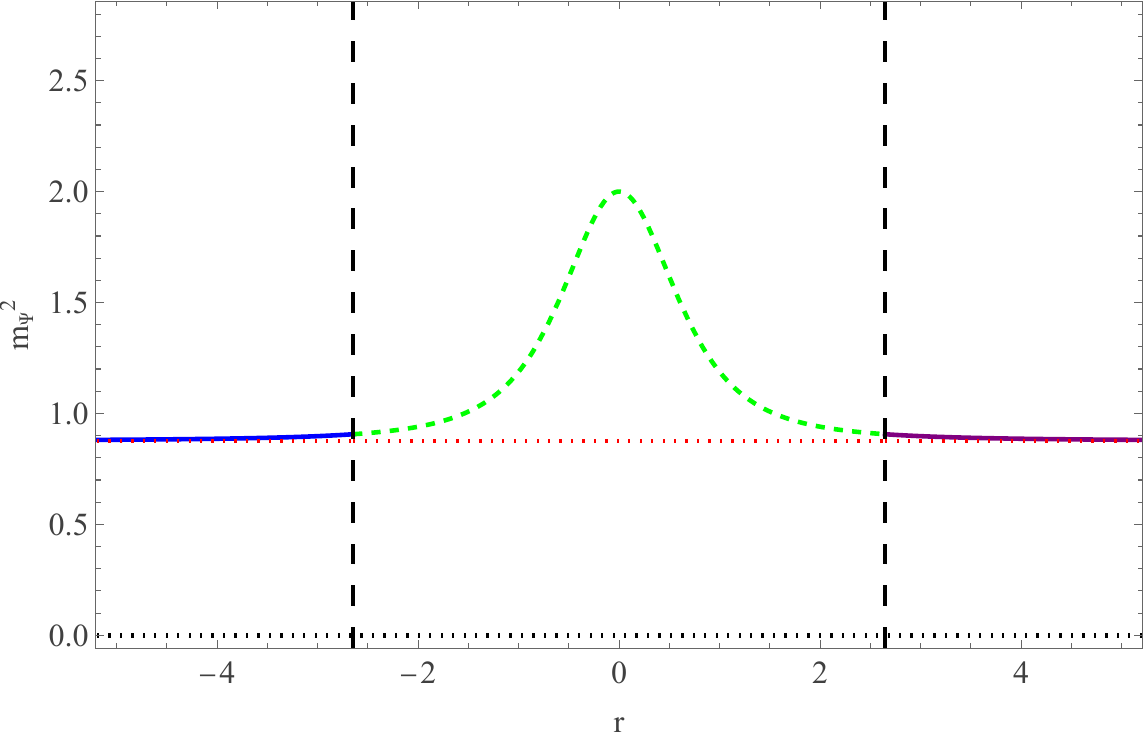}
\caption{}
\end{subfigure}

\caption{Graphical representation of $m_{\Psi}^2$ as a function of the radial coordinate $r$. The black dashed vertical lines denote the horizons, the red dotted horizontal line indicates the asymptotic value at radial infinity, the blue and purple curves represent the mass outside the horizon (always positive), and the green dashed curve corresponds to the mass inside the horizons. Panels a), b), and c) depict the models $f_R=1+a_2 r^2$, $f_R=1+a_{\Sigma}\Sigma$, and $f_R=1+2a_R R$, respectively.}
\label{msmods}
\end{figure}

For the Starobinsky model,
\begin{equation}
f(R)=R+a_R R^2,
\end{equation}
the viability conditions take a particularly simple form. One finds
\begin{equation}
f_R(R)=\frac{df}{dR}=1+2a_R R,
\qquad
f_{RR}(R)=\frac{d^2f}{dR^2}=2a_R .
\end{equation}
Hence, the requirements $f_R>0$ and $f_{RR}>0$ imply
\begin{equation}
1+2 a_R R>0 \;\;\Longleftrightarrow\;\; R>-\frac{1}{2a_R}.
\end{equation}
Moreover, substituting these expressions into Eq.~(\ref{ms}) yields the scalaron mass
\begin{eqnarray}
&&m_{\Psi}^{2}
=\frac{1}{8} \Bigg[\frac{4}{l^2} \left(\frac{a^2}{a^2+r^2}-3\right)+\frac{4 a^2
   M}{\left(a^2+r^2\right)^2}+\frac{1}{a_R}\Bigg],\nonumber\\
&&m_{\Psi}^{2}(r\rightarrow\infty)\approx \frac{1}{8} \left(\frac{1}{a_R}-\frac{12}{l^2}\right)+\frac{a^2}{2 l^2
   r^2}+O\left(r^{-3}\right). 
\end{eqnarray}
We can choose the parameters such that the effective mass of the scalaron is positive throughout the entire spacetime. We illustrate the case $M =2 ,a =1 ,l =2 ,a_R=0.1$ in Fig.~\ref{msmods}~c).
\subsection{Model \texorpdfstring{$R=0$}{}}
We can use the results from the previous subsection to verify the case where $R=0$. Therefore, at $R=0$ the viability conditions reduce to $f_R(0)=1>0$ and $f_{RR}(0)=2a_R>0$, implying $a_R>0$. Moreover, Eq.~(\ref{ms}) gives
\begin{equation}
m_\Psi^2\Big|_{R=0}=\frac{1}{8 a_R},
\end{equation}
so that $m_\Psi^2>0$ provided $a_R>0$ (i.e., no tachyonic scalaron at $R=0$).

\section{Energy Conditions}  \label{EN}
A central requirement for assessing the physical viability of $f(R)$ gravity models is the explicit verification of the energy conditions. In these theories, the field equations can be reinterpreted through an effective stress-energy tensor, necessitating a detailed analysis of the resulting fluid components. Such an investigation is relevant for BB geometries, which are characterized by their ability to bridge the transition between regular BHs and WHs. Since these configurations typically remove the singularities found in GR at the cost of violating classical energy conditions \cite{Simpson:2018tsi,Lobo:2020ffi,Bronnikov:2018vbs}, it is essential to determine if these violations persist when the gravitational action is modified.

Consequently, we analyze the null (NEC), weak (WEC), strong (SEC), and dominant (DEC) energy conditions by treating the source as an effective anisotropic fluid. With the energy density $\rho$, radial pressure $p_r$, and tangential pressure $p_t$ as the components of the stress-energy tensor, the conditions are defined as \cite{Visser:1995cc}:
\begin{eqnarray}
&&NEC_{1,2}\Longleftrightarrow \rho+p_{r,t}\geq 0,\label{Econd1}\\
&&WEC_{1,2}\Longleftrightarrow \rho\geq 0\ \ \text{and}\ \ \rho+p_{r,t}\geq 0,\label{EcondW}\\
&&SEC_{1,2}\Longleftrightarrow \rho+p_{r,t}\geq 0,
\qquad
SEC_3\Longleftrightarrow \rho+p_r+p_t\geq 0,\label{Econd2}\\
&&DEC_{1,2}\Longleftrightarrow \rho\geq 0\ \ \text{and}\ \ \rho-|p_{r,t}|\geq 0
\ \Longleftrightarrow\ (\rho\pm p_{r,t}\geq 0),\label{Econd3}\\
&&WEC_3=DEC_3\Longleftrightarrow \rho\geq 0.\label{Econd4}
\end{eqnarray}
The subscripts $1$ and $2$ denote relations involving the radial and tangential components, while $3$ corresponds to the energy density alone or the specific trace-related combinations.

The functions $\{\rho, p_r, p_t\}$ are derived from the mixed components of the effective stress-energy tensor. In the exterior regions or where $A(r)>0$, we adopt the standard identification
\begin{equation}
T^\mu_{\ \nu}=\mathrm{diag}\!\left[\rho,\,-p_r,\,-p_t,\,-p_t\right].
\end{equation}
However, in regions where $A(r)<0$, the causal roles of the time and radial coordinates are interchanged. To maintain a consistent physical interpretation in these regions, the tensor components are defined as
\begin{equation}
T^\mu_{\ \nu}=\mathrm{diag}\!\left[-p_r,\,\rho,\,-p_t,\,-p_t\right].
\end{equation}
We now examine how each energy condition behaves for the models considered in this work.

\subsection{Model \texorpdfstring{$f_R = 1 + a_2 r^2$}{}}

\begin{figure}
    \centering
    \includegraphics[width=.5\linewidth]{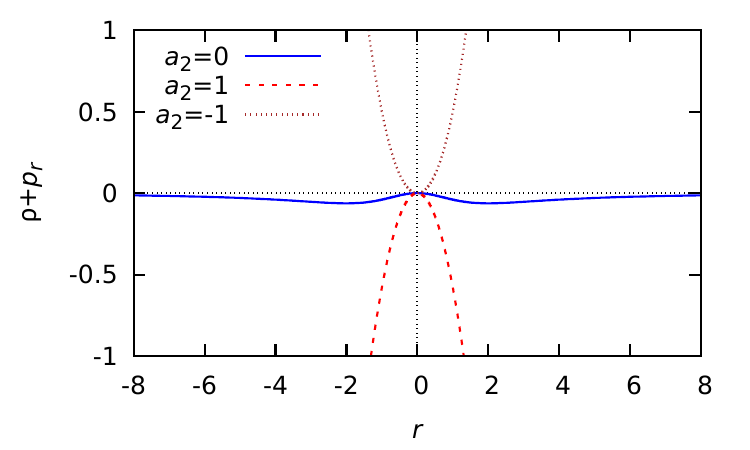}\hspace{-0.5cm}
    \includegraphics[width=.5\linewidth]{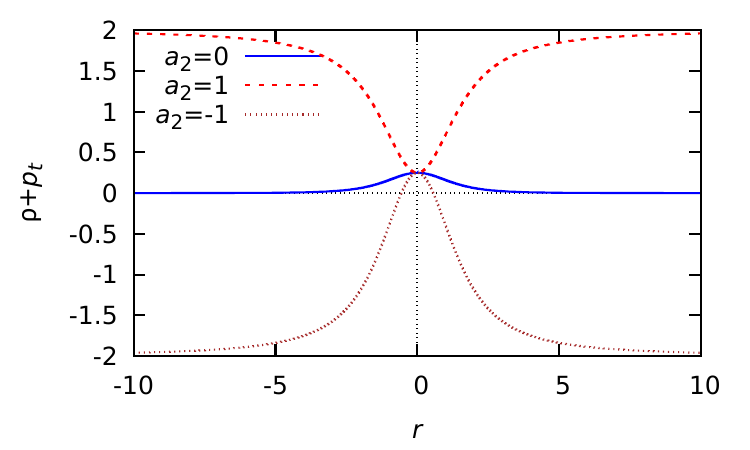}
    \includegraphics[width=.5\linewidth]{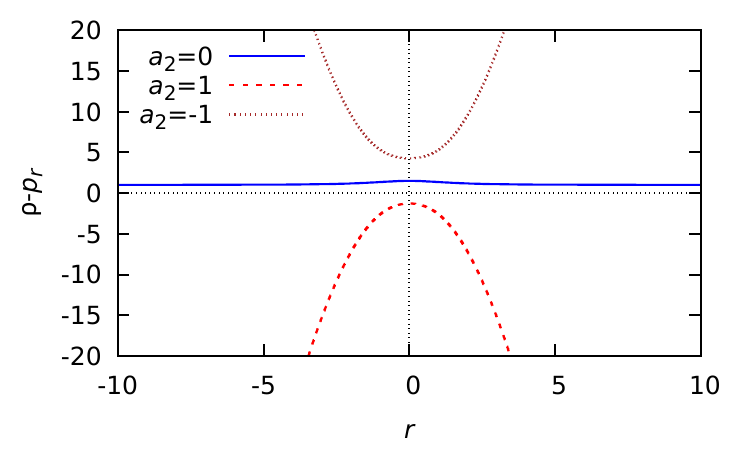}\hspace{-0.5cm}
    \includegraphics[width=.5\linewidth]{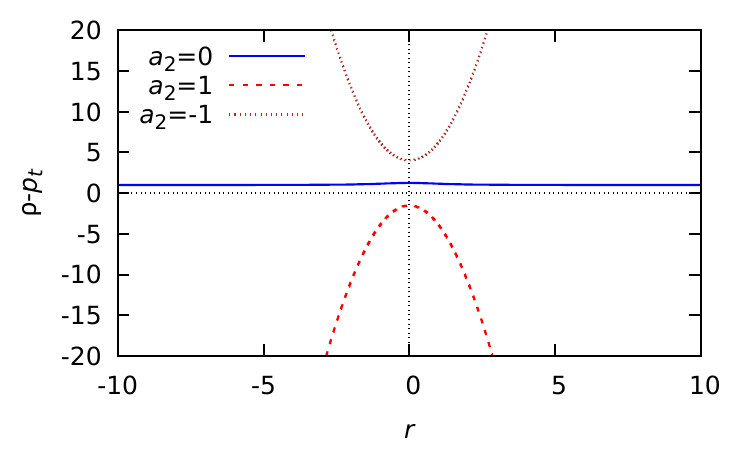}
      \includegraphics[width=.5\linewidth]{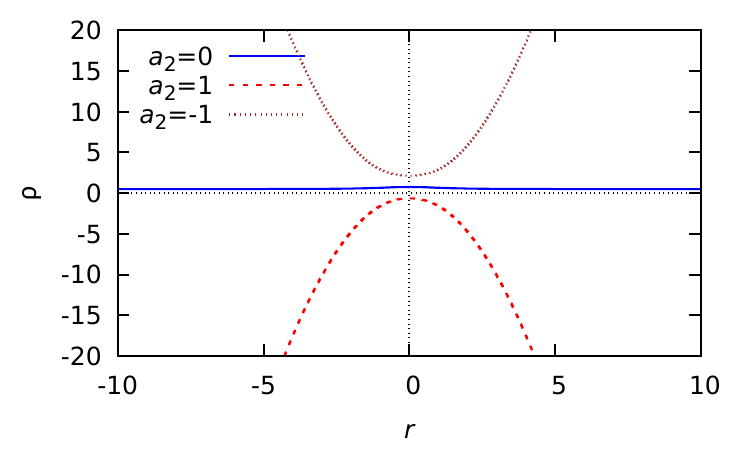}\hspace{-0.5cm}
      \includegraphics[width=.5\linewidth]{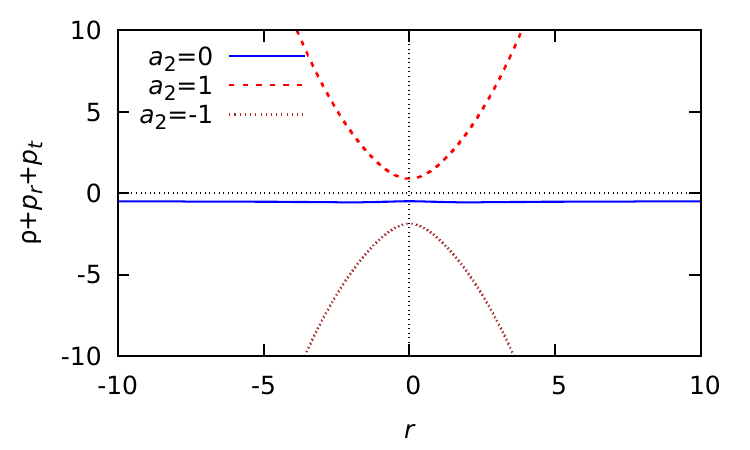}
    \caption{Combinations of the stress-energy tensor components for the $f_R =1+ a_2 r^2$ model as a function of the radial coordinate, with $a=l = 2$ and $M = 1$, for different values of $a_2$. The combinations are invariant under $r \to -r$, due to the symmetry of $f_{R}$. For the chosen set of parameters, the horizon is located at the same point as
the throat, at $r = 0$.}
    \label{fig:EC_model1}
\end{figure}

Let us now analyze the energy conditions for the case where $f_R=1+a_2 r^2$. For regions where $A(r) > 0$, we have:
\begin{eqnarray}
    \rho + p_r &=&-\frac{a^2 \left(a^2-l^2
   M+r^2\right)}{l^2 \left(a^2+r^2\right)^2}-\frac{a_2 \left(2 a^2+r^2\right) \left(a^2+2 r^2\right) \left(a^2-l^2 M+r^2\right)}{l^2 \left(a^2+r^2\right)^2}, \\
    \rho + p_t &=&\frac{a^2 M}{\left(a^2+r^2\right)^2}+\frac{a_2 M \left(3 a^2 r^2+2 r^4\right)}{\left(a^2+r^2\right)^2},\\
\rho + p_r + p_t &=&\frac{a^2 M}{\left(a^2+r^2\right)^2}-\frac{\frac{a^2}{a^2+r^2}+2}{l^2}+\frac{a_2 \left(a^6+a^4 \left(3 r^2-l^2 M\right)+4 a^2 r^4+a^2
   \left(a^2+r^2\right)^2 \ln \left(a^2+r^2\right)+2 r^6\right)}{l^2 \left(a^2+r^2\right)^2},\nonumber\\
\end{eqnarray}
\begin{eqnarray}
\rho - p_r &=&\frac{\frac{a^2}{a^2+r^2}+4}{l^2}+\frac{a^2 M}{\left(a^2+r^2\right)^2}\\&-&\frac{a_2 \left(4 a^6+a^4 \left(13 r^2-4 l^2 M\right)+a^2 \left(15
   r^4-11 l^2 M r^2\right)+2 a^2 \left(a^2+r^2\right)^2 \ln \left(a^2+r^2\right)-6 l^2 M r^4+6 r^6\right)}{l^2 \left(a^2+r^2\right)^2},\nonumber\\
    \rho - p_t &=&\frac{4}{l^2} +\frac{a^2 M}{\left(a^2+r^2\right)^2}\\&-&\frac{a_2 \left(6 a^6+a^4 \left(20 r^2-6 l^2 M\right)+a^2 \left(22 r^4-13 l^2 M r^2\right)+2 a^2
   \left(a^2+r^2\right)^2 \ln \left(a^2+r^2\right)-6 l^2 M r^4+8 r^6\right)}{l^2 \left(a^2+r^2\right)^2},\nonumber\\
\rho &=&\frac{2}{l^2}+\frac{a^2 M}{\left(a^2+r^2\right)^2}\\
&-&\frac{a_2 \left(3 a^6+a^4 \left(10 r^2-3 l^2 M\right)+a^2 \left(11 r^4-8 l^2 M r^2\right)+a^2
   \left(a^2+r^2\right)^2 \ln \left(a^2+r^2\right)-4 l^2 M r^4+4 r^6\right)}{l^2 \left(a^2+r^2\right)^2} .\nonumber
\end{eqnarray}
In regions where $A(r)<0$, we have
\begin{eqnarray}
    \rho + p_r &=&\frac{a^2 \left(a^2-l^2
   M+r^2\right)}{l^2 \left(a^2+r^2\right)^2}+\frac{a_2 \left(2 a^2+r^2\right) \left(a^2+2 r^2\right) \left(a^2-l^2 M+r^2\right)}{l^2 \left(a^2+r^2\right)^2} ,\\
    \rho + p_t &=&\frac{a^2}{l^2 \left(a^2+r^2\right)}+\frac{a_2 \left(2 a^4+a^2 \left(5 r^2-2 l^2 M\right)+2 r^4\right)}{l^2 \left(a^2+r^2\right)}, \\
\rho + p_r + p_t &=&-\frac{a^4+a^2 \left(l^2 M+3 r^2\right)+2 r^4}{l^2
   \left(a^2+r^2\right)^2}\\
   &+&\frac{a_2 \left(5 a^6+a^4 \left(17 r^2-5 l^2 M\right)+2 a^2 r^2 \left(9 r^2-5 l^2 M\right)+a^2 \left(a^2+r^2\right)^2 \ln
   \left(a^2+r^2\right)-4 l^2 M r^4+6 r^6\right)}{l^2 \left(a^2+r^2\right)^2},\nonumber
\end{eqnarray}
\begin{eqnarray}
\rho - p_r &=& \frac{\frac{a^2}{a^2+r^2}+4}{l^2}+\frac{a^2 M}{\left(a^2+r^2\right)^2}\\
&-&\frac{a_2 \left(4 a^6+a^4 \left(13 r^2-4 l^2 M\right)+a^2 \left(15
   r^4-11 l^2 M r^2\right)+2 a^2 \left(a^2+r^2\right)^2 \ln \left(a^2+r^2\right)-6 l^2 M r^4+6 r^6\right)}{l^2 \left(a^2+r^2\right)^2},\nonumber\\  
\rho - p_t &=&\frac{\frac{a^2}{a^2+r^2}+4}{l^2}-\frac{a_2 \left(4 a^4+a^2 \left(9 r^2-4 l^2 M\right)+2 a^2 \left(a^2+r^2\right) \ln
   \left(a^2+r^2\right)-4 l^2 M r^2+6 r^4\right)}{l^2 \left(a^2+r^2\right)},\\
\rho &=& \frac{\frac{a^2}{a^2+r^2}+2}{l^2}-\frac{a_2 \left(a^4+a^2 \left(2 r^2-l^2 M\right)+a^2 \left(a^2+r^2\right) \ln \left(a^2+r^2\right)-2
   l^2 M r^2+2 r^4\right)}{l^2 \left(a^2+r^2\right)}.
\end{eqnarray}
 The analytic expressions for this model are quite complex, and it is more feasible to study these combinations through plots, Fig. \ref{fig:EC_model1}. What we observe is that, depending on the sign of $a_{2}$, some of the combinations become strictly positive. The issue is that not all combinations are positive for the same sign of $a_{2}$. For instance, $\rho+p_{r}$ is positive if $a_{2}<0$, whereas $\rho+p_{t}$ is negative over most of the spacetime. Thus, the energy conditions will always be violated in some regions; it is not possible to satisfy all of them simultaneously for a single sign of $a_{2}$.

\subsection{Model \texorpdfstring{$f_R = 1 + a_\Sigma \Sigma$}{}}

In this case, where $f_R =1+a_\Sigma \Sigma$, we have, outside any possible horizon, $A(r)>0$, that the following combinations for the functions $\rho$, $p_r$, and $p_t$ are
\begin{eqnarray}
    \rho + p_r &=&-\frac{a^2 \left(a^2-l^2 M+r^2\right)}{l^2
   \left(a^2+r^2\right)^2} -\frac{2 a^2 a_\Sigma  \left(a^2-l^2 M+r^2\right)}{l^2 \left(a^2+r^2\right)^{3/2}},\\
\rho + p_t &=&\frac{a^2 M}{\left(a^2+r^2\right)^2}+\frac{a_\Sigma  M}{\sqrt{a^2+r^2}} ,\\
\rho + p_r + p_t &=&-\frac{\frac{a^2}{a^2+r^2}+2}{l^2}+\frac{a^2
   M}{\left(a^2+r^2\right)^2}+ \frac{2 a^2 a_\Sigma \left(-6 a^2+l^2 M-6 r^2\right)}{3 l^2 \left(a^2+r^2\right)^{3/2}},
\end{eqnarray}
\begin{eqnarray}
\rho - p_r &=&\frac{4}{l^2}+\frac{a^2}{l^2 \left(a^2+r^2\right)}+\frac{a^2 M}{\left(a^2+r^2\right)^2}+ \frac{2 a_\Sigma\left(9 a^4+a^2 \left(4 l^2 M+9 r^2\right)+3 l^2 M r^2\right)}{3 l^2 \left(a^2+r^2\right)^{3/2}} ,\\
\rho - p_t &=&\frac{a^2
   M}{\left(a^2+r^2\right)^2}+\frac{4}{l^2}+a_\Sigma  \left(\frac{4 a^2}{l^2 \sqrt{a^2+r^2}}+\frac{8 a^2 M}{3 \left(a^2+r^2\right)^{3/2}}+\frac{M}{\sqrt{a^2+r^2}}\right),\\
\rho &=&\frac{a^2
   M}{\left(a^2+r^2\right)^2}+\frac{2}{l^2} +a_\Sigma  \left(\frac{2 a^2}{l^2 \sqrt{a^2+r^2}}+\frac{4 a^2 M}{3 \left(a^2+r^2\right)^{3/2}}+\frac{M}{\sqrt{a^2+r^2}}\right).
\end{eqnarray}

Whereas, for $A < 0$, we have:
\begin{eqnarray}
       \rho + p_r &=& \frac{a^2
   M}{\left(a^2+r^2\right)^2}+\frac{2}{l^2}+a_\Sigma \left(\frac{2 a^2}{l^2 \sqrt{a^2+r^2}}+\frac{4 a^2 M}{3 \left(a^2+r^2\right)^{3/2}}+\frac{M}{\sqrt{a^2+r^2}}\right),\\
  \rho +  p_t &=& \frac{a^2}{l^2 \left(a^2+r^2\right)}+\frac{a_\Sigma \left(2 a^4-a^2 l^2 M+2 a^2 r^2+l^2 M r^2\right)}{l^2 \left(a^2+r^2\right)^{3/2}},\\
   \rho + p_r + p_t &=&-\frac{a^2+2 r^2}{l^2 \left(a^2+r^2\right)}-\frac{a^2 M}{\left(a^2+r^2\right)^2}-\frac{10 a^2 a_\Sigma M}{3 \left(a^2+r^2\right)^{3/2}} ,
\end{eqnarray}
\begin{eqnarray}
\rho - p_r &=&\frac{4}{l^2}+ \frac{a^2}{l^2 \left(a^2+r^2\right)}+\frac{a^2 M}{\left(a^2+r^2\right)^2}+ \frac{2 a_\Sigma\left(9 a^4+a^2 \left(4 l^2 M+9 r^2\right)+3 l^2 M r^2\right)}{3 l^2 \left(a^2+r^2\right)^{3/2}},\\
\rho - p_t &=& \frac{\frac{a^2}{a^2+r^2}+4}{l^2}+\frac{a_\Sigma \left(18 a^4+a^2 \left(5 l^2 M+18 r^2\right)+3 l^2 M r^2\right)}{3 l^2
   \left(a^2+r^2\right)^{3/2}},\\
\rho &=& \frac{\frac{a^2}{a^2+r^2}+2}{l^2}+\frac{a_\Sigma \left(12 a^4+a^2 \left(l^2 M+12 r^2\right)+3 l^2 M r^2\right)}{3 l^2
   \left(a^2+r^2\right)^{3/2}}.
\end{eqnarray}
The combinations for the model $f_{R}=1+a_{\Sigma},\Sigma$ are simpler than in the previous case. From the analytic expressions, we see that, outside the event horizon, we can ensure the positivity of $\rho+p_{t}$, $\rho-p_{r}$, $\rho-p_{t}$, and $\rho$, provided $a_{\Sigma}>0$. Inside the horizon, we can ensure the positivity of almost all combinations if $2a^{2}>l^{2}M$, except for $\rho+p_{r}+p_{t}$, still with $a_{\Sigma}>0$. Thus, inside the horizon all energy conditions can be satisfied except the strong energy condition.

\subsection{Model \texorpdfstring{$f(R) = R + a_R R^2$}{}}
Let us now consider the corrections to the energy conditions for the model $f(R) = R+ a_R R^2$. In regions where $A(r) > 0 $, we have
\begin{eqnarray}
    \rho + p_r &=&-\frac{a^2 \left(a^2-l^2 M+r^2\right)}{l^2 \left(a^2+r^2\right)^2}-\frac{4 a^2 a_R \left(a^2-l^2 M+r^2\right) \left(4 a^4+a^2 \left(3 l^2
   M+r^2\right)-20 l^2 M r^2-3 r^4\right)}{l^4 \left(a^2+r^2\right)^4},\\
    \rho + p_t &=& \frac{a^2 M}{\left(a^2+r^2\right)^2}+\frac{4 a^2 a_R M \left(2 a^4+a^2 \left(7 r^2-l^2 M\right)+4 l^2 M r^2+5 r^4\right)}{l^2
   \left(a^2+r^2\right)^4},\\
    \rho + p_r + p_t &=&-\frac{a^2}{l^2 \left(a^2+r^2\right)}+\frac{a^2 M}{\left(a^2+r^2\right)^2}\nonumber\\
    &+&a_R \left(\frac{12 a^2}{l^4 \left(a^2+r^2\right)}+\frac{52 a^2
   M}{l^2 \left(a^2+r^2\right)^2}-\frac{14 a^4}{l^4 \left(a^2+r^2\right)^2}-\frac{36 a^4 M}{l^2 \left(a^2+r^2\right)^3}-\frac{6 a^4
   M^2}{\left(a^2+r^2\right)^4}-\frac{6}{l^4}\right)+\frac{1}{l^2},
\end{eqnarray}
\begin{eqnarray}
        \rho - p_r &=&-\frac{2}{l^2}+ \frac{a^2}{l^2 \left(a^2+r^2\right)}+\frac{a^2 M}{\left(a^2+r^2\right)^2}\nonumber\\
    &+& {a_R} \left(-\frac{12 a^2}{l^4 \left(a^2+r^2\right)}+\frac{4 a^2
   M}{l^2 \left(a^2+r^2\right)^2}-\frac{16 a^2 M \left(a^2+3 l^2 M\right)}{l^2 \left(a^2+r^2\right)^3}+\frac{64 a^4
   M^2}{\left(a^2+r^2\right)^4}+\frac{12}{l^4}\right) ,\\
    \rho - p_t &=& -\frac{2}{l^2}+\frac{a^2 M}{\left(a^2+r^2\right)^2}+4 a_R \left(\frac{4 a^2 M^2 \left(2 a^2-9 r^2\right)}{\left(a^2+r^2\right)^4}+\frac{3-\frac{7
   a^4}{\left(a^2+r^2\right)^2}}{l^4}+\frac{13 a^2 M r^2-4 a^4 M}{l^2 \left(a^2+r^2\right)^3}\right), \\
    \rho &=&-\frac{1}{l^2}+\frac{a^2 M}{\left(a^2+r^2\right)^2}+2 a_R \left(-\frac{2 a^2 M \left(a^2-9 r^2\right)}{l^2 \left(a^2+r^2\right)^3}+\frac{a^2 M^2 \left(7
   a^2-32 r^2\right)}{\left(a^2+r^2\right)^4}+\frac{3-\frac{7 a^4}{\left(a^2+r^2\right)^2}}{l^4}\right).
\end{eqnarray}
\begin{figure}
    \centering
    \includegraphics[width=.5\linewidth]{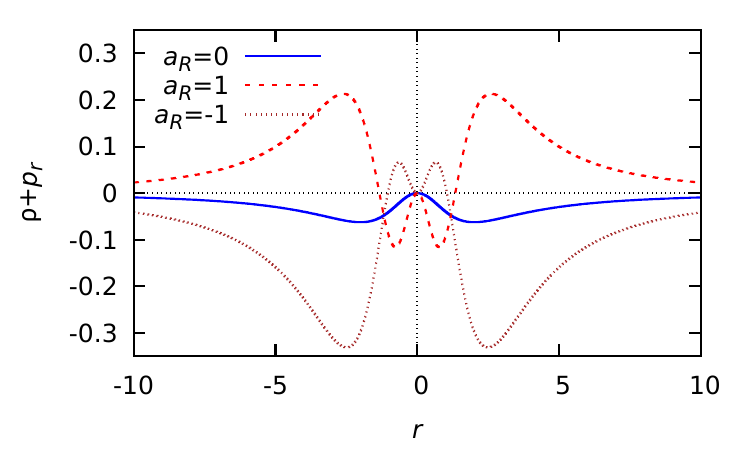}\hspace{-.5cm}    \includegraphics[width=.5\linewidth]{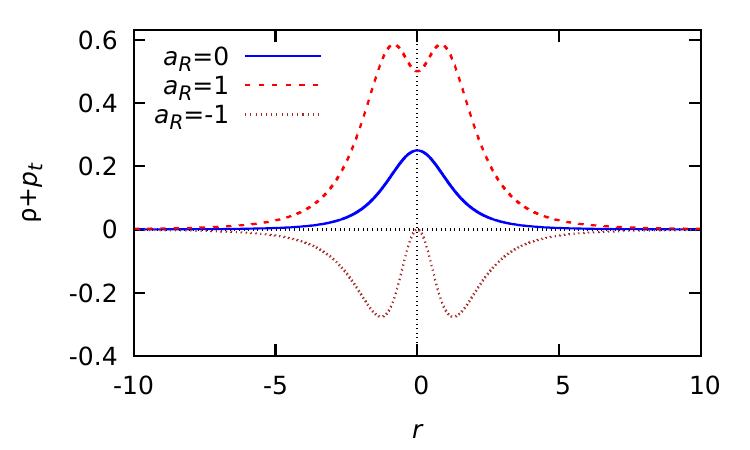}    \includegraphics[width=.5\linewidth]{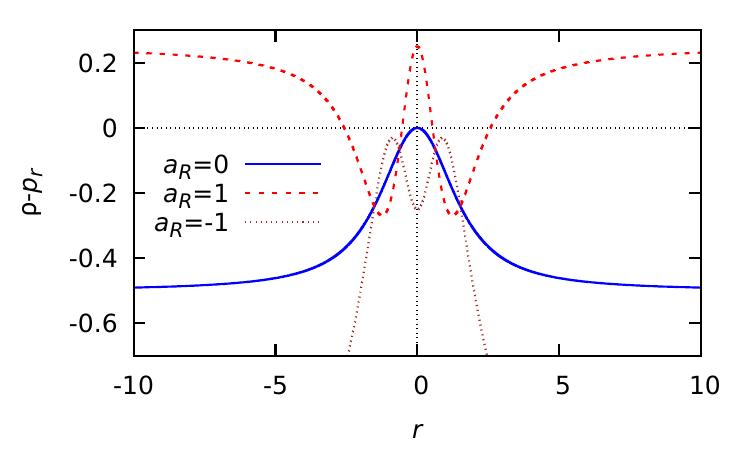}\hspace{-.5cm}      \includegraphics[width=.5\linewidth]{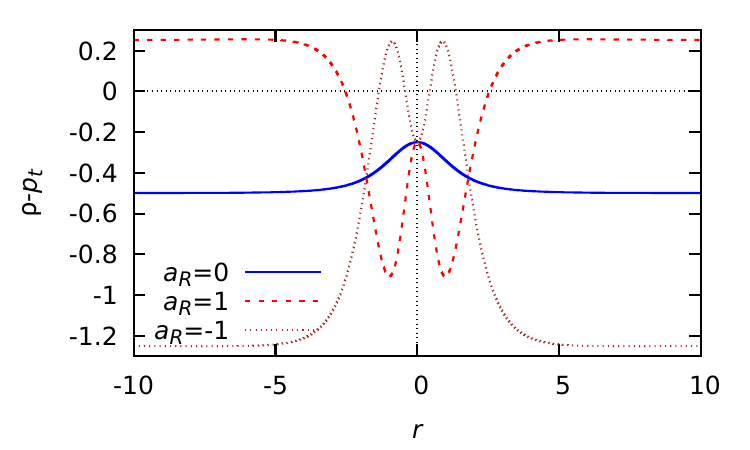}      \includegraphics[width=.5\linewidth]{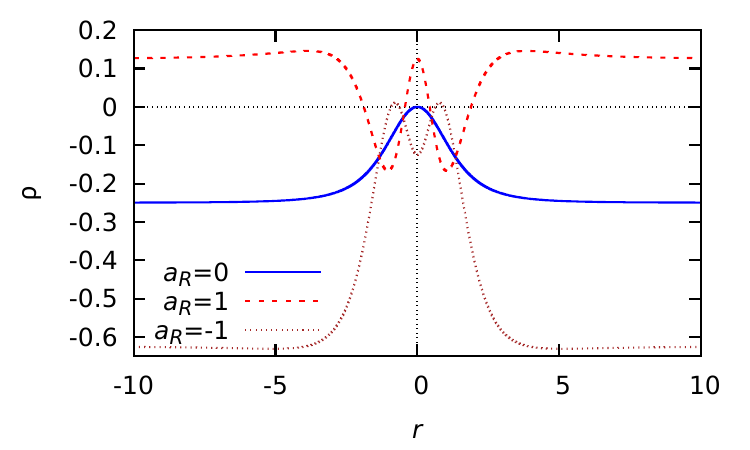}\hspace{-.5cm}     \includegraphics[width=.5\linewidth]{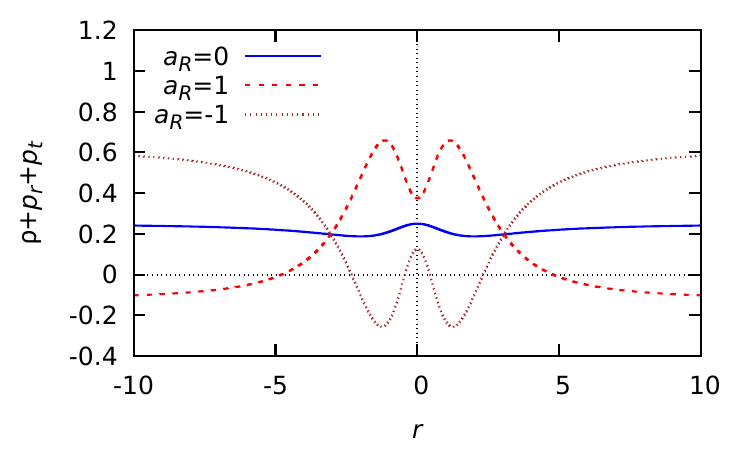}
    \caption{Combination of the stress-energy tensor components for the model $f(R) = R+a_R R^2$, as a function of the radial coordinate, for $M = 1$, $a=l=2$, and different values of $a_R$. For the chosen set of parameters, the horizon is located at the same point as the throat, at $r=0$. Depending on the value of the constant $a_R$, the regions where these combinations are positive or negative may change.}
    \label{fig:EC_model3}
\end{figure}

In regions where $A(r) < 0$, we have
\begin{eqnarray}
 \rho + p_r &=&\frac{a^2 \left(a^2-l^2 M+r^2\right)}{l^2 \left(a^2+r^2\right)^2}+\frac{4 a^2 a_R \left(a^2-l^2 M+r^2\right) \left(4 a^4+a^2 \left(3 l^2
   M+r^2\right)-20 l^2 M r^2-3 r^4\right)}{l^4 \left(a^2+r^2\right)^4} ,\\
 \rho + p_t &=&\frac{a^2}{l^2 \left(a^2+r^2\right)}\nonumber\\
 &+&\frac{4 a^2 a_R \left(4 a^6+a^4 l^2 M-4 a^2 l^4 M^2-2 r^4 \left(a^2+6 l^2 M\right)+r^2 \left(5 a^4-11
   a^2 l^2 M+24 l^4 M^2\right)-3 r^6\right)}{l^4 \left(a^2+r^2\right)^4} ,\\
\rho + p_r + p_t &=&\frac{1}{l^2}+\frac{a^2}{l^2 \left(a^2+r^2\right)}-\frac{a^2 M}{\left(a^2+r^2\right)^2}\nonumber\\
&+&a_R \left(-\frac{12 a^2}{l^4 \left(a^2+r^2\right)}+\frac{4 a^2 M \left(23 a^2+40 l^2 M\right)}{l^2 \left(a^2+r^2\right)^3}+\frac{42 a^4-84a^2l^2M}{l^4
   \left(a^2+r^2\right)^2}-\frac{190 a^4 M^2}{\left(a^2+r^2\right)^4}-\frac{6}{l^4}\right),
\end{eqnarray}
\begin{eqnarray}
     \rho -p_r &=&-\frac{2}{l^2}+\frac{a^2}{l^2 \left(a^2+r^2\right)}+\frac{a^2 M}{\left(a^2+r^2\right)^2}\nonumber\\
 &+&a_R \left(-\frac{12 a^2}{l^4 \left(a^2+r^2\right)}+\frac{4 a^2
   M}{l^2 \left(a^2+r^2\right)^2}-\frac{16 a^2 M \left(a^2+3 l^2 M\right)}{l^2 \left(a^2+r^2\right)^3}+\frac{64 a^4
   M^2}{\left(a^2+r^2\right)^4}+\frac{12}{l^4}\right) ,\\   
\rho - p_t &=&-\frac{2}{l^2}+ \frac{a^2}{l^2 \left(a^2+r^2\right)}+a_R \left(\frac{12}{l^4}-\frac{12 a^2}{l^4 \left(a^2+r^2\right)}-\frac{16 a^2 M}{l^2
   \left(a^2+r^2\right)^2}-\frac{4 a^2 M \left(a^2+16 l^2 M\right)}{l^2 \left(a^2+r^2\right)^3}+\frac{84 a^4
   M^2}{\left(a^2+r^2\right)^4}\right),\nonumber\\\\
\rho &=&-\frac{r^2}{l^2 \left(a^2+r^2\right)}\\
&+&\frac{2 a_R \left(a^2 r^4 \left(7 a^2-16 l^2 M\right)+6 a^2 r^6+a^4 \left(l^2 M-2 a^2\right)^2+4 a^2 r^2 \left(2 a^4-5 a^2 l^2 M+2 l^4
   M^2\right)+3 r^8\right)}{l^4 \left(a^2+r^2\right)^4}.\nonumber
\end{eqnarray}
As we can see, the combinations for this model are more involved than in the previous cases and therefore an analytic analysis is unfeasible. Thus, we will examine the combinations through plots Fig.~\ref{fig:EC_model3}. As we can see, almost all combinations, for both $a_{R}>0$ and $a_{R}<0$, exhibit negative regions, so the energy conditions for this model are violated even more than in the previous models.

\subsection{Model \texorpdfstring{$R=0$}{}}            
Let us now analyze the energy conditions for the final model, $R=0$. In the regions where $A > 0$, we have
\begin{eqnarray}
    \rho + p_r &=&\frac{a \left(4 a M \cos \left(\frac{\pi -2 \tan ^{-1}\left(\frac{r}{a}\right)}{\sqrt{2}}\right)+\sqrt{2} Q \sin \left(\frac{\pi -2 \tan
   ^{-1}\left(\frac{r}{a}\right)}{\sqrt{2}}\right)\right)}{4 \left(a^2+r^2\right)^2} ,\\
    \rho + p_t &=& \frac{\left(8 a^2 M-3 Q r\right) \cos \left(\frac{\pi -2 \tan ^{-1}\left(\frac{r}{a}\right)}{\sqrt{2}}\right)+2 \sqrt{2} a (3 M r+Q) \sin
   \left(\frac{\pi -2 \tan ^{-1}\left(\frac{r}{a}\right)}{\sqrt{2}}\right)}{4 \left(a^2+r^2\right)^2},\\
    \rho + p_r + p_t &=& \frac{\left(4 a^2 M-Q r\right) \cos \left(\frac{\pi -2 \tan ^{-1}\left(\frac{r}{a}\right)}{\sqrt{2}}\right)+\sqrt{2} a (2 M r+Q) \sin
   \left(\frac{\pi -2 \tan ^{-1}\left(\frac{r}{a}\right)}{\sqrt{2}}\right)}{2 \left(a^2+r^2\right)^2},\\
    \rho - p_r &=& \frac{\left(4 a^2 M-2 Q r\right) \cos \left(\frac{\pi -2 \tan ^{-1}\left(\frac{r}{a}\right)}{\sqrt{2}}\right)+\sqrt{2} a (4 M r+Q) \sin
   \left(\frac{\pi -2 \tan ^{-1}\left(\frac{r}{a}\right)}{\sqrt{2}}\right)}{4 \left(a^2+r^2\right)^2},\\
    \rho - p_t &=& \frac{r \left(Q \cos \left(\frac{\pi -2 \tan ^{-1}\left(\frac{r}{a}\right)}{\sqrt{2}}\right)-2 \sqrt{2} a M \sin \left(\frac{\pi -2 \tan
   ^{-1}\left(\frac{r}{a}\right)}{\sqrt{2}}\right)\right)}{4 \left(a^2+r^2\right)^2} ,\\
    \rho &=&\frac{\left(4 a^2 M-Q r\right) \cos \left(\frac{\pi -2 \tan ^{-1}\left(\frac{r}{a}\right)}{\sqrt{2}}\right)+\sqrt{2} a (2 M r+Q) \sin
   \left(\frac{\pi -2 \tan ^{-1}\left(\frac{r}{a}\right)}{\sqrt{2}}\right)}{4 \left(a^2+r^2\right)^2} .
\end{eqnarray}

\begin{figure}
    \centering
    \includegraphics[width=.6\linewidth]{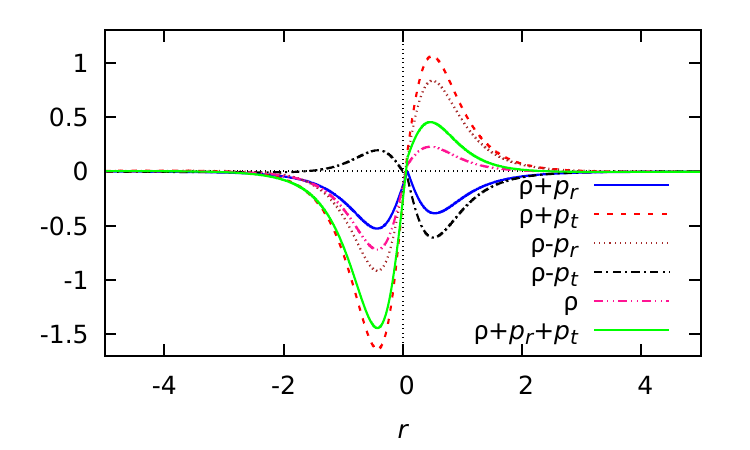}
    \caption{Combination of the stress-energy tensor components for the model $R=0 $, as a function of the radial coordinate, for $M = a = q = 1$.}
    \label{fig:EC_model4}
\end{figure}

In regions where $A<0$, we have
\begin{eqnarray}
    \rho + p_r &=& -\frac{a \left(4 a M \cos \left(\frac{\pi -2 \tan ^{-1}\left(\frac{r}{a}\right)}{\sqrt{2}}\right)+\sqrt{2} Q \sin \left(\frac{\pi -2 \tan
   ^{-1}\left(\frac{r}{a}\right)}{\sqrt{2}}\right)\right)}{4 \left(a^2+r^2\right)^2},\\
\rho + p_t &=&\frac{\left(4 a^2 M-3 Q r\right) \cos \left(\frac{\pi -2 \tan ^{-1}\left(\frac{r}{a}\right)}{\sqrt{2}}\right)+\sqrt{2} a (6 M r+Q) \sin
   \left(\frac{\pi -2 \tan ^{-1}\left(\frac{r}{a}\right)}{\sqrt{2}}\right)}{4 \left(a^2+r^2\right)^2},\\
\rho + p_r + p_t &=&\frac{r \left(2 \sqrt{2} a M \sin \left(\frac{\pi -2 \tan ^{-1}\left(\frac{r}{a}\right)}{\sqrt{2}}\right)-Q \cos \left(\frac{\pi -2 \tan
   ^{-1}\left(\frac{r}{a}\right)}{\sqrt{2}}\right)\right)}{2 \left(a^2+r^2\right)^2},\\
 \rho - p_r &=&\frac{\left(4 a^2 M-2 Q r\right) \cos \left(\frac{\pi -2 \tan ^{-1}\left(\frac{r}{a}\right)}{\sqrt{2}}\right)+\sqrt{2} a (4 M r+Q) \sin
   \left(\frac{\pi -2 \tan ^{-1}\left(\frac{r}{a}\right)}{\sqrt{2}}\right)}{4 \left(a^2+r^2\right)^2} ,\\
\rho - p_t &=& \frac{\left(Q r-4 a^2 M\right) \cos \left(\frac{\pi -2 \tan ^{-1}\left(\frac{r}{a}\right)}{\sqrt{2}}\right)-\sqrt{2} a (2 M r+Q) \sin
   \left(\frac{\pi -2 \tan ^{-1}\left(\frac{r}{a}\right)}{\sqrt{2}}\right)}{4 \left(a^2+r^2\right)^2},\\
    \rho &=&\frac{r \left(2 \sqrt{2} a M \sin \left(\frac{\pi -2 \tan ^{-1}\left(\frac{r}{a}\right)}{\sqrt{2}}\right)-Q \cos \left(\frac{\pi -2 \tan
   ^{-1}\left(\frac{r}{a}\right)}{\sqrt{2}}\right)\right)}{4 \left(a^2+r^2\right)^2} .
\end{eqnarray}
Since the curvature scalar vanishes for this model, the energy conditions do not depend on the constant $a_{R}$, as expected. The behavior of the combinations of the stress-energy tensor components is shown in Fig.~\ref{fig:EC_model4}. We observe that $\rho+p_{r}$ is always negative, so the NEC is always violated and, consequently, the other energy conditions are violated as well.

\section{Conclusions and discussion} \label{conclusions}
In this work, we focused on investigating BB solutions in $2+1$ dimensions within the framework of $f(R)$ gravity. We examined which types of matter sources are capable of sustaining such exotic geometries in modified gravity theories. One of our main approaches was to determine which $f(R)$ models and source fields can generate a regularized BTZ solution. A second approach was to explore the possibility of constructing a solution with vanishing curvature scalar for a given $f(R)$ model.  

Our results show that it is indeed possible to obtain a regularized BTZ solution in $f(R)$ gravity if one considers a NED source coupled to a partially phantom scalar field. This is a noteworthy result when compared to GR, where the regularized BTZ solution necessarily requires a scalar field that is always phantom.  

In the first case, we adopted the function $f_{R}=1+a_{n}r^{n}$, which represents a generalization inspired by the forms that naturally arise in regular BH solutions in $f(R)$ and $f(G)$ theories, where typically $f_{R}=1+a_{1}r$. Since the resulting functions become highly involved in the general case, we focused on the illustrative example $n=2$. We found that the functions associated with the source fields are more complicated than in GR, with the corrections induced by $f(R)$ gravity being controlled by the constant $a_{2}$. A notable feature of this example is the symmetry under $r\to -r$, which implies that the source-field functions are also symmetric. We showed that a canonical scalar field can be obtained if $a_{2}$ takes negative values and satisfies $-a^{2}a_{2}>1/2$. However, the $f(R)$ model satisfies the viability conditions $f_{R}>0$ and $f_{RR}>0$ throughout the spacetime only if $a_{2}>0$. Consequently, the viability requirements enforce a scalar field that is partially phantom. Moreover, the condition $a_{2}>0$ implies that, in some regions, at least asymptotically, the scalaron mass becomes negative. Concerning the energy conditions, we found that negative values of $a_{2}$ allow the inequality associated with $\mathrm{NEC}_{1}$, which is always satisfied in GR, to be fulfilled. However, this choice simultaneously implies that other inequalities are violated even more strongly than in the GR case. Thus, in a global sense, the energy conditions remain violated, even though some of the inequalities that were violated in GR can now be at least partially satisfied.

Another proposal for $f_{R}$ arises from applying the Simpson--Visser regularization procedure directly to the function $f_{R}$ that appears in regular BH models. In fact, this construction can be made even more general by considering $f_{R}=1+a_{\Sigma}\Sigma^{n}$. As a simple illustrative example, we analyzed the case $n=1$ and derived the corresponding source field functions. In general, the functions obtained for $f_{R}=1+a_{\Sigma}\Sigma$ are simpler than those arising in the case $f_{R}=1+a_{2}r^{2}$. The associated scalar field can be everywhere canonical if $a_{\Sigma}$ takes negative values and satisfies the condition $-a_{\Sigma}a>1/2$. In the electromagnetic sector, it is not possible to invert $F(r)$ due to the complexity of this function; however, in some cases the Lagrangian $L(F)$ is not multivalued. In this approach, we were able to obtain an analytic expression for $f(R)$, which is not always feasible, especially in $3+1$ dimensions. When analyzing the viability conditions of this $f(R)$ model, we found that $f_{R}>0$ and $f_{RR}>0$ are satisfied if $a_{\Sigma}>0$. However, this choice implies that the scalaron mass becomes negative in some regions and that the scalar field associated with the solution is at least partially phantom. Concerning the energy conditions, we observed that the inequality associated with the strong energy condition is the one for which positivity can never be guaranteed, either inside or outside the event horizon. By contrast, inside the horizon it is possible to ensure the positivity of all remaining combinations. Thus, apart from the strong energy condition, the energy conditions can be satisfied inside the horizon if $a_{\Sigma}>0$. In this sense, the choice $a_{\Sigma}>0$ contributes simultaneously to satisfying both the energy conditions (with the exception of the strong one) and the viability conditions of the $f(R)$ theory.

We also considered the case of the Starobinsky model, in which the form of $f(R)$, and consequently $f_{R}$. In this scenario, the functions associated with the matter sector are considerably more involved than in the previous cases. The electromagnetic invariant $F(r)$ develops both maxima and minima, which renders the Lagrangian $L(F)$ multivalued and prevents us from writing it in closed analytic form. For the scalar field, it is never possible to enforce a condition that makes it canonical throughout the entire spacetime: if $a_{R}>0$, the scalar field can be canonical in asymptotic regions but becomes phantom in more central regions, whereas for $a_{R}<0$ this behavior is reversed. When examining the viability conditions of the $f(R)$ theory, we found that the model can easily satisfy $f_{R}>0$ and $f_{RR}>0$, which ensures that the scalaron mass remains positive, provided that $a_{R}>0$. From the perspective of the energy conditions, however, the coupling to $f(R)$ gravity leads to an even stronger violation of the energy conditions, in many cases more severe than in GR. The advantage of this coupling is that some inequalities that were always violated in GR can now be satisfied, but only in restricted regions of the spacetime.

As a final case, we considered the situation in which $R=0$. In this scenario, we were able to solve Einstein’s equations and, by imposing that the solution should behave asymptotically as the Einstein--CIM model, we related the integration constants to the mass and charge of the BH. To some extent, the resulting geometry can be regarded as a BB solution, since it exhibits both a throat and horizons. However, we found that in asymptotic regions with $r>0$ the metric function satisfies $A(r)<0$, so that the solution actually describes an inverted BH, with the static region located inside the event horizon. Depending on the values of the parameters, this horizon may lie either in the region $r>0$ or $r<0$. We also determined the matter sources for this model and showed once again that a combination of a phantom scalar field (in this case always phantom) with NED can sustain the solution. In the electromagnetic sector, the function $F(r)$ is highly nontrivial, displaying several maxima and minima, and the corresponding Lagrangian $L(F)$ exhibits multiple branches, thus being a multivalued function. The viability conditions are easily satisfied, since $f_{R}=1>0$ and $f_{RR}=2a_{R}>0$, with the scalaron mass also being positive, $m_{\psi}^{2}=1/(6a_{R})>0$. When analyzing the energy conditions, however, we found that all inequalities are violated in some region of the spacetime, and in fact some of them are violated everywhere. Consequently, this solution violates all the standard energy conditions.

Regarding the phantom behavior of the scalar field, we observe that in the case $f_{R}=1+a_{2}r^{2}$ the scalar field is always phantom if $a_{2}>0$, which is precisely the sign required by the viability conditions of the $f(R)$ theory. Thus, the viability of the $f(R)$ model demands that the scalar field be phantom in this case. For the model $f_{R}=1+a_{\Sigma}\Sigma$, the scalar field is also always phantom if $a_{\Sigma}>0$, which again corresponds to the requirement imposed by the viability conditions of this $f(R)$ theory. In the Starobinsky case, the constant $a_{R}>0$ leads to a scalar field that behaves as phantom in regions closer to the center and canonical in asymptotically distant regions, thus corresponding to a partially phantom scalar field. If $a_{R}<0$, this behavior is reversed, with the field being phantom at large distances and canonical near $r=0$. However, we focus on the case $a_{R}>0$, since this choice is also required by the viability conditions of the $f(R)$ model. Finally, in the case where we impose $R=0$, the scalar field is always phantom.

In this way, we have shown that it is indeed possible to obtain BB solutions in $2+1$ dimensions within $f(R)$ gravity. These solutions can be described by a combination of NED and a scalar field. Depending on the specific solution and on the choice of parameters, the scalar field can be always phantom, partially phantom, or canonical. However, scalar fields that are everywhere canonical typically violate the viability conditions of the $f(R)$ theory. In order to satisfy the viability requirements, one is therefore led to impose that the scalar field be phantom. The source-field models are more involved and more nonlinear than in the GR case. From the perspective of the energy conditions, it is only possible to guarantee that they are partially satisfied in some cases, while in others they are always violated in at least some region of the spacetime.

While BB solutions in 2+1 dimensions have been recently obtained in $f(R, R_{\mu\nu})$ gravity \cite{Alencar:2024nxi}, our work provides a distinct perspective by focusing on the $f(R)$ sector. A key advantage of our approach is the explicit construction of the matter source using the $H(P)$ formalism. This allows us to avoid the multi-valued branches often encountered in NED models and to establish a direct relationship between the 'inverted' causal structure and the nonlinear electromagnetic field. Furthermore, our solutions with $R=0$ represent a specific class of analytical geometries that differ from the regularized BTZ seeds found in higher-curvature theories.

Beyond the comparison with specific higher-curvature models, the novelty of this work lies in providing a systematic construction of field sources for $(2+1)-$dimensional BBs within the widely applicable $f(R)$ gravity framework. While most BB seeds are historically derived in the context of GR, our results demonstrate that these geometries are robust enough to be consistently generalized to modified gravity sectors without losing their essential causal features. By identifying the precise coupling between the scalar field and nonlinear electrodynamics required to support such spacetimes, we bridge the gap between abstract geometric constructions and consistent field-theoretical sources. This establishes $f(R)$ gravity as a viable and mathematically transparent arena for exploring the properties of regular BHs and wormholes in lower dimensions.

As perspectives for future works it would be of interest to investigate whether the qualitative features found here persist for other classes of BB geometries in $2+1$ dimensions and for different functional forms of $f(R)$. In addition, a more detailed study of perturbative stability, quasinormal modes, and possible observational signatures, such as shadow properties or lensing effects,in the context of $2+1$-dimensional BBs in modified gravity would provide further insight into the physical relevance of these solutions. 

\section*{Acknowledgments}
M.S. thanks Conselho Nacional de Desenvolvimento Cient\'ifico e Tecnol\'ogico - CNPq, Brazil, CNPQ/PDE 200218/2025-5, for financial support. MER thanks Conselho Nacional de Desenvolvimento Cient\'ifico e Tecnol\'ogico - CNPq, Brazil, for partial financial support.


%

\end{document}